\documentclass[journal]{IEEEtran}
\usepackage{amsmath,amsfonts}
\usepackage{algorithmic}
\usepackage{algorithm}
\usepackage{textgreek}
\usepackage{array}
\usepackage[caption=false,font=normalsize,labelfont=sf,textfont=sf]{subfig}
\usepackage{textcomp}
\usepackage{stfloats}
\usepackage{url}
\usepackage{verbatim}
\usepackage{graphicx}
\usepackage{cite}
\usepackage{enumitem}
\usepackage{booktabs}
\usepackage{float}
\usepackage{multirow}
\usepackage{longtable}
\usepackage{xr}

\begin{document}

\title{Feature Encoding in VAE-based Audio Decoders: Effects of Input, Depth and Distribution}
\author{Louis McCallum and Mick Grierson
\thanks{The authors are with the Creative Computing Institute, 
University of the Arts London, London, UK 
(e-mail: l.mccallum@arts.ac.uk; m.grierson@arts.ac.uk).}
\thanks{Code for the activation extraction and analysis 
pipeline is available at https://github.com/Louismac/rave-activations}
}

\markboth{IEEE/ACM Transactions on Audio, Speech, and Language Processing }{McCallum and Grierson: Where Pitch and Tempo Live}

\noindent\fbox{%
  \parbox{\textwidth}{%
    \textbf{Preprint.} This manuscript has been accepted for publishing in IEEE Transactions on Audio, Speech and Language Processing (TASLP). 
    Licensed under CC BY 4.0. \url{https://creativecommons.org/licenses/by/4.0/}
  }%
}

\maketitle
\thispagestyle{empty}

\begin{abstract}
Neural audio synthesis models like the Realtime Audio Variational autoEncoder (RAVE) achieve impressive generation quality, yet how their internal representations encode musical features remains poorly understood. We present a systematic layer-wise and cross-layer cluster analysis of RAVE decoder activations across three models trained on different musical domains, tested with four stimulus types. We then evaluate architectural generalization with a general purpose EnCodec model. For RAVE, we find that synthetic stimuli are encoded well across models and audio features (pitch $|\rho|$=0.45, 5.1$\times$ the null, BPM $|\rho|$ = 0.76, 8.6$\times$ the null). These results are reduced but still substantively apparent when using natural audio (mean across features $|\rho|$=0.25, 2.8$\times$ the null). Natural audio sees a stronger encoding when nonlinear probes are used (mean across features $R^2$=0.56, 18$\times$ the null, +0.152 nonlinear gain over the linear probe $R^2$). Encoding strength varies throughout the layers of the decoder and an increased ability to joint-encode in the middle layers is seen across all audio features ($\beta_2$ all negative, $p<0.05$). The general purpose EnCodec decoder also sees similar strong synthetic responses across audio features, similar nonlinear gains for natural audio joint encoding and similar depth profiles. We find the best cross-layer cluster improves the strength ($r=0.65$, $p=0.006$) and prevalence ($r=0.75$, $p=0.001$) of BPM encoding when compared against the best whole layers within the same section, with no effect for joint encoding. These findings advance the interpretability of neural audio models and inform targeted control strategies for neural synthesis.
\end{abstract}

\begin{IEEEkeywords} Audio synthesis, clustering methods, feature extraction, neural networks, variational autoencoders (VAEs)\end{IEEEkeywords}

\section{Introduction}

\IEEEPARstart{A}{dvances} in neural audio synthesis have enabled real-time, high-fidelity music generation through models like RAVE \cite{caillon2021rave}, EnCodec \cite{defossez2022highfidelity}, and SoundStream \cite{zeghidour2021soundstream}. These variational autoencoder (VAE)-based architectures learn compressed latent representations that can reconstruct audio with perceptual quality rivaling traditional synthesis methods. However, understanding what musical features these models encode, and where in their architecture this encoding occurs, remains largely unexplored.

This knowledge gap matters for both scientific understanding and practical application. From a scientific perspective, neural audio models are increasingly deployed in creative and research contexts, yet we lack basic knowledge of whether pitch, tempo, or timbre occupy identifiable locations within these architectures or whether they are encoded in distributed, entangled representations that resist decomposition. From a practical standpoint, if we can identify which decoder layers or neuron populations represent specific musical features, this understanding could inform the design of more targeted control interfaces for neural synthesizers, structured pruning strategies, and diagnostic tools for model debugging.

Existing work has examined learning constraints in latent space \cite{engel2018latent} and controllability \cite{huang2018improved}, but layer-wise activation analysis of the kind common in computer vision \cite{zeiler2014visualizing} has received limited attention in audio. Studies of self-supervised speech models such as Wav2Vec \cite{baevski2020wav2vec} have revealed hierarchical acoustic feature learning, with early layers capturing spectral patterns and later layers encoding semantic content. Layer-wise probing of models including HuBERT and WavLM has demonstrated that pitch, tempo, and energy encoding varies systematically with network depth \cite{pasad2024self, chiu2026largescaleprobinganalysisspeakerspecific}. Whether similar hierarchical organization exists in generative audio models remains an open question.

A methodological challenge in this domain is distinguishing genuine feature encoding from chance correlation. We address this through permutation testing \cite{good2005permutation}, establishing empirical null distributions for each model-dataset combination and defining feature prevalence (the percentage of neurons exceeding the permutation-derived 95th percentile threshold, referred to by \% $>p_{95}$) as a complement to conventional encoding strength (mean $|\rho|$). This dual-metric approach reveals aspects of encoding that neither measure captures alone: encoding strength reflects how intensely neurons track a feature, while prevalence reflects how broadly that tracking is distributed across the neural population. We further use nonlinear probes to establish the abilities of neuron subset to jointly encode these features (measured by $R^2$). Where valid, results are compared to EnCodec \cite{defossez2022highfidelity} to investigate whether the findings are specific to RAVE's variational design.

We address:

\textbf{RQ1:} How do linear and nonlinear probes compare across synthetic and natural audio?

\textbf{RQ2:} How does feature encoding vary with depth in the network?

\textbf{RQ3:} Can cross-layer clustering identify neuron groups that encode features more effectively than individual layers?

We contribute:
\begin{enumerate}
 \item 
Systematic layer-wise analysis across three models $\times$ four datasets using encoding strength, permutation-validated feature prevalence and nonlinear probes.
\item Evidence that RAVE decoders encode pitch ($|\rho|$ = 0.45, 5.1$\times$ the null) and BPM ($|\rho|$ = 0.76, 8.6$\times$ the null) strongly, linearly and broadly throughout layers when presented with synthetic stimuli and that this is maintained for natural audio even when spectral features are also tested (mean across features $|\rho|$ = 0.25, 2.8$\times$ the null). Natural audio sees a stronger encoding when nonlinear probes are used (mean across features $R^2$=0.56, 18$\times$ the null, +0.152 nonlinear gain over the linear probe $R^2$).
\item We see some complete and some feature specific patterns in responses of layers through the model when considering depth. Evidence of all four features ($\beta_2$ all negative, $p<0.05$) shows non-monotonic joint encoding with mid-network peaks. Three of the four features (pitch, centroid, and bandwidth $\beta_2$ all positive, $p<0.05$) show per-neuron concentration toward late layers.
\item We find the best cross-layer cluster improves the strength ($r=0.65$, $p=0.006$) and prevalence ($r=0.75$, $p=0.001$) of BPM encoding when compared against the best whole layers within the same section. We find no improvement for joint-encoding power.
\item A cross-architecture comparison showing the findings are not specific to RAVE's variational design.
\end{enumerate}

Our analysis proceeds in sequential steps, each building upon the results of previous work. We examine feature encoding on a per-layer basis, investigating how strongly and where musical features are represented across RAVE’s 21 decoder layers, and whether hierarchical patterns are shared across models or domain-specific. We then extend this to cross-layer analysis, using K-means clustering to identify neuron populations spanning multiple layers that collectively encode features more coherently than any single layer. Together, these analyses provide a detailed picture of how one of the most commonly used neural audio synthesis architectures internally represents musical structure and timbral information.

\section{Related Work}
Our work combines neural audio synthesis and network interpretability. We review prior work in each area, identify the gap our analysis addresses, and position our methodological contributions.

\subsection{Neural Audio Synthesis and Generative Models}

\subsubsection{Neural audio codecs}
Recent advances in neural audio compression have produced high-quality generative models that learn compact latent representations of audio. SoundStream~\cite{zeghidour2021soundstream} introduced residual vector quantization for efficient audio coding, while EnCodec~\cite{defossez2022highfidelity} extended this approach with improved perceptual quality. RAVE~\cite{caillon2021rave}, the focus of our analysis, employs variational autoencoders with adversarial training to achieve real-time audio synthesis with latency suitable for musical applications. More recent generative architectures have moved in two directions with transformer-based autoregressive models such as MusicGen~\cite{copet2023simple} and AudioLM~\cite{borsos2023audiolmlanguagemodelingapproach} that operate over discrete codebook tokens, and diffusion-based synthesizers such as AudioLDM~\cite{liu2023audioldm} that introduce iterative denoising. Despite this proliferation, the convolutional VAE family remains the dominant choice for real-time musical applications where latency constraints preclude autoregressive or diffusion-based alternatives. Across all of these architectures the question of what and where their internal representations encode has received limited empirical treatment.

\subsubsection{Audio feature encoding}
Understanding what neural networks learn about audio has been explored primarily through probing classifiers~\cite{alain2016understanding}, in which a supervised model is trained to predict an acoustic or linguistic feature from frozen layer activations. Studies of Wav2Vec~\cite{baevski2020wav2vec} and related self-supervised models reveal hierarchical acoustic feature learning, with early layers capturing spectral patterns and late layers encoding semantic content. Pasad et al.~\cite{pasad2024selfsupervised} extend this paradigm to lexical content, while Chiu et al.~\cite{chiu2026largescaleprobinganalysisspeakerspecific} conduct layer-wise probing of HuBERT, WavLM, and Wav2Vec 2.0 for pitch, tempo, and energy, finding a consistent three-stage hierarchy across discriminative speech models. 

This body of work has, however, two limitations relevant to our setting. First, it focuses almost exclusively on encoders trained for discriminative or self-supervised objectives whereas the analogous question for generative decoders has not been systematically addressed. Second, reliance on trained probes raises capacity confounds~\cite{hewitt2019designing} in that a sufficiently expressive probe can extract features that are computable from the representation rather than genuinely encoded in it. Our analysis addresses both gaps by extending layer-wise probing to a generative decoder and complementing trained probes with rank-correlation and permutation-based prevalence measures that are robust to capacity confounds.

\subsection{Neural Network Interpretability and Mechanistic Understanding}

Neural network interpretability encompasses a broad range of approaches that can be organized along several dimensions: passive versus active methods, the type of explanation produced (examples, attribution, hidden semantics, or rules), and the scope of interpretation (local, semi-local, or global)~\cite{zhang2021survey}. Our work occupies a specific position in this landscape as we perform passive, post-hoc analysis aimed at uncovering hidden semantics at a global scope that characterizes what neuron populations across entire decoder hierarchies encode, rather than explaining individual predictions or modifying the architecture for interpretability.

\begin{figure*}
    \centering
    \includegraphics[width=0.8\linewidth]{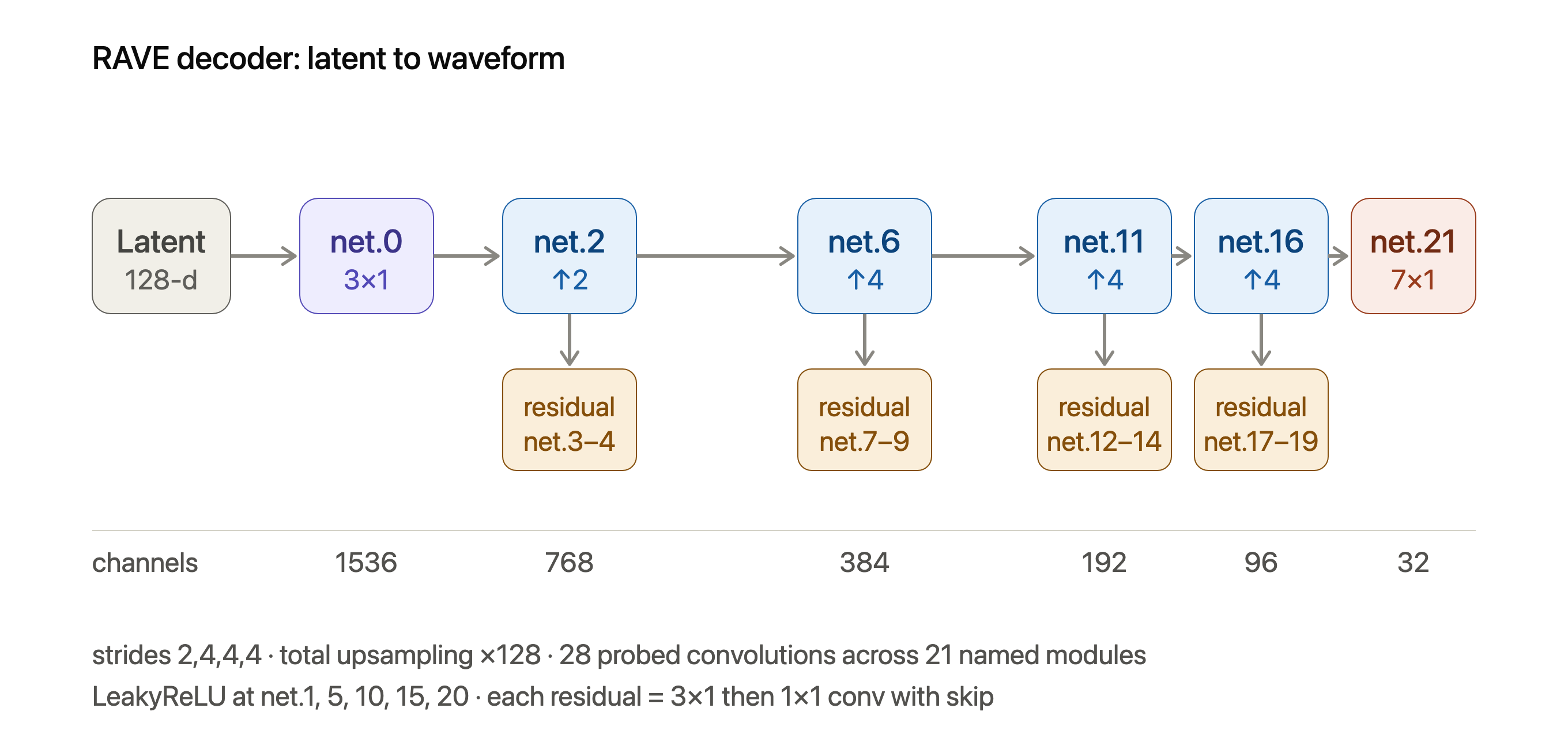}
    \caption{Diagram of RAVE Decoder architecture}
    \label{fig:rave}
\end{figure*}

\subsubsection{Activation-based interpretation}
Visualising and manipulating neuron activations has become a cornerstone of interpretability research. Feature visualization~\cite{olah2017feature} demonstrates that scaling activations amplifies or suppresses feature detection in vision models. Deep visualization~\cite{yosinski2015understanding} reveals neuron selectivity through activation maximization. Network dissection~\cite{bau2017network} quantifies interpretability by measuring neuron-concept alignment against a labelled visual concept dataset. These methods share an assumption that meaningful features align with individual neurons or simple linear combinations thereof and subsequent work has shown this to be only partially correct as features are frequently encoded in superposition across overlapping neuron populations~\cite{elhage2022superposition}. This motivates the more distributed, prevalence-based view of feature encoding we adopt here.

\subsubsection{Attribution methods}
A complementary family of interpretability methods assigns credit to input features for a model's output. Layer-wise Relevance Propagation (LRP)~\cite{bach2015lrp} decomposes predictions by propagating relevance scores backward through the network. DeepLIFT~\cite{shrikumar2017learning} compares activations to a reference to propagate activation differences, while Integrated Gradients~\cite{sundararajan2017axiomatic} accumulates gradients along a path from a baseline input to provide axiomatic attribution. SHAP~\cite{lundberg2017shap} unifies several attribution approaches under a game-theoretic framework. Although powerful for explaining individual predictions, attribution methods are ill-suited to the question we address as they identify which inputs drive an output, not which internal representations carry specific information about a feature. For generative models in particular, where the meaningful "output" is a high-dimensional waveform rather than a class label, the attribution framing maps poorly onto questions of musical feature encoding, motivating our focus on activation-based rather than attribution-based analysis.

\subsubsection{Circuits and mechanistic interpretability}
Cammarata et al.~\cite{cammarata2020thread} seek to reverse-engineer neural networks by identifying computational subgraphs that implement specific functions. They emphasize hierarchical feature composition in which early layers detect simple patterns that later layers compose into complex representations. Subsequent work in this tradition has identified phenomena that offer increasingly fine-grained accounts of how trained networks represent and manipulate information. This includes induction heads in language models~\cite{olsson2022incontext}, polysemantic neurons and feature superposition~\cite{elhage2022superposition}, and sparse-autoencoder-based decomposition of activations into monosemantic features~\cite{bricken2023monosemanticity}.

Two observations from this line of work bear directly on our setting. First, the mechanistic interpretability program has focused on language models and, to a lesser extent, vision models. Audio generative models remain largely unexamined despite their growing deployment. Second, the central tension between localized (per-neuron) and distributed (cross-neuron, cross-layer) encoding remains an open question for any new architecture. Our prevalence-based and cluster-based analyses address this tension directly for RAVE. Rather than presupposing that features are localized to individual neurons or that they are diffusely distributed, we measure where on this spectrum each feature actually sits.

\subsubsection{Neuron clustering}
Grouping neurons by functional similarity has proven effective for understanding network organization. Raghu et al.~\cite{raghu2017svcca} introduced SVCCA for comparing layer representations via canonical correlation analysis, revealing that similar layers cluster together. B\"{a}uerle et al.~\cite{bäuerle2022neuralactivationpatternsnaps} find groups of images that are treated similarly by extracting similar activation profiles in the activation space of a neural network layer. A common limitation of these approaches is that clustering is typically performed within individual layers, leaving open the question of whether functionally-related neurons spanning across multiple layers might cohere into more informative groupings than any single layer provides. We address this directly through cross-layer clustering.

\subsection{Summary of the Gap}
Having covered the relevant literature, we summarize the gap our work addresses. First, layer-wise probing of audio representations has been concentrated on discriminative speech encoders and comparable analysis of generative musical decoders is largely absent. Second, existing audio interpretability work rarely separates encoding strength (how intensely individual units track a feature) from encoding prevalence (how broadly that tracking is distributed across the neural population). This conflates two distinguishable properties that the superposition and circuits literatures suggest can dissociate. Third, neuron-clustering analyses have typically been confined to within-layer groupings, leaving cross-layer functional organization in generative architectures unexamined. Our contributions of a three-metric layer-wise analysis and cross-layer clustering address this gap for the RAVE family, with a methodological framework with potential to be portable to other architectures.

\section{Method}

To balance focused internal validation against reliable musical properties and externally valid real world sounds, we present a probe analysis on three models and four datasets (three naturally occurring and one synthetic). This means that we can draw conclusions for 12 combinations and draw insights from differences and similarities between combinations. This helps us move towards broad and reliable claims about RAVE models' encoding and decoding of music in more general terms. We further compare these results with EnCodec, a pretrained general purpose decoder \cite{defossez2022highfidelity}.

\subsection{Training Datasets}

Models tend to converge best on homogeneous datasets that contain a focused set of timbres. However, within this they can successfully train on sets of sounds that are polyphonic, multi-instrumental, a-rhythmic and even nonmusical (e.g. Foley sounds). As such, it is highly challenging to find a single RAVE model with enough timbral breadth to draw fully generalizable conclusions about universal modeling performance. 

We collected three datasets of distinct categories of musical recordings to train three RAVE models from scratch. The strings (N=740) and drums (N=759) datasets consist of samples from a commercial online library \cite{splice}. The strings were selected to allow for a good distribution of key and tempo during downstream balancing and include solo and ensemble. Drums were also selected to allow good distribution of tempo during downstream balancing and are spread across a variety of contemporary genres including rock, pop, hip hop, jazz and jungle. The vocals dataset (N=486) is the full catalog of a commercially available pop singer, where vocals have been separated using Meta's Demucs algorithm \cite{rouard2022hybrid}. 

\subsection{Model}

RAVE employs a variational autoencoder architecture in which the encoder compresses raw audio into a low dimensional latent space and the decoder reconstructs the waveform through a series of upsampling and convolutional stages (see Fig~\ref{fig:rave}). This comparatively simple architecture (when compared to other models such as transformers) makes the extraction and analysis of activations straightforward. 

We trained three models on the datasets described above (strings, drums, vocals). Each uses an identical decoder architecture and only the training data differs. The models are mono and have a 44.1 kHz sample rate with 128 latent dimensions. The datasets are within the good practice lengths (minimum 2--3 hours), and the models were trained for at least 1.5M steps using a standard RAVE V2 approach \cite{caillon2022streamable}.

Treating each convolution as a separate layer yields 28 probed units across 21 named modules. Activations are captured after weight normalization and nonlinearity but before residual addition, isolating each layer's transformation. 

\begin{table*}[t]
\centering
\caption{Acoustic feature distributions per dataset after per-feature balanced 
sampling. All cells use N=500 samples, stratified to span the full feature range. 
Exclusions for inappropriate feature / data combinations apply}
\label{tab:dataset_features}
\renewcommand{\arraystretch}{1.15}
\setlength{\tabcolsep}{4pt}
\small
\begin{tabular}{l | r r r | r r r | r r r | r r r}
\toprule
& \multicolumn{3}{c|}{\textbf{Pitch (Hz)}} 
  & \multicolumn{3}{c|}{\textbf{BPM}} 
  & \multicolumn{3}{c|}{\textbf{Spectral Centroid (Hz)}} 
  & \multicolumn{3}{c}{\textbf{Spectral Bandwidth (Hz)}} \\
\textbf{Dataset} 
  & Min & Max & Mean ($\sigma$) 
  & Min & Max & Mean ($\sigma$)
  & Min & Max & Mean ($\sigma$)
  & Min & Max & Mean ($\sigma$) \\
\midrule
Strings 
  & 80 & 788 & 293 (200)
  & 70 & 140 & 104 (21)
  & 306 & 7357 & 1976 (1123)
  & 969 & 6082 & 2601 (813) \\
Drums 
  & --- & --- & ---
  & 60 & 175 & 115 (34)
  & 652 & 7902 & 3961 (1711)
  & 948 & 6074 & 3553 (1105) \\
Vocals 
  & 80 & 761 & 284 (126)
  & 65 & 123 & 90 (17)
  & 918 & 7905 & 3938 (995)
  & 1803 & 6608 & 3799 (779) \\
Synthetic 
  & 80 & 800 & 313 (200)
  & 60 & 180 & 120 (35)
  & --- & --- & ---
  & --- & --- & ---\\

\bottomrule
\end{tabular}
\end{table*}

\subsection{Model Inputs}

\subsubsection{Sampling}
To assess how well the models encode audio features, we constructed four sets of audio (three from the training datasets and one synthesized) and forward-passed them through each model while collecting activations. Each audio segment is paired with feature labels that are combined with the activations for subsequent analysis. Valid results require sets that are balanced in the distribution of each feature, drawn from diverse source audio, and sufficiently large to yield adequate statistical confidence.

As a baseline representing the most reduced signal with minimal confounding variation, we generated two synthetic datasets, one varying in pitch and the other in tempo. The pitch set comprises of sine tones at equally spaced interval and the tempo set is made of pulse trains. 


To determine whether the observed effects extend to more complex, naturally occurring audio, and to assess the influence of in- and out-of-distribution inputs, we constructed representative, balanced subsets of the data used to train the three models. 

For drums, the audio is divided into four second chunks and randomly sampled to obtain a uniform distribution across tempo, prioritizing source-file diversity. For strings, the same procedure is applied with uniform sampling across both tempo and pitch. For vocals, samples are selected to favor high RMS energy (unlikely to be silence) and samples are randomly sampled to obtain a uniform distribution across tempo and pitch, again prioritizing source-file diversity. This results in 500 BPM balanced and 500 pitch balanced samples where appropriate.

\subsubsection{Features}
We choose four features for this analysis, namely, pitch, tempo (BPM), spectral centroid and spectral bandwidth. Pitch is a fundamental perceptual attribute of musical audio that contributes to melody and harmony \cite{mcdermott2008music}. Tempo captures rhythmic structure that unfolds over time rather than within a single frame, testing whether the model encodes temporally-extended properties as well as instantaneous ones. Spectral centroid provides the perceived brightness of a sound and is a good correlate of timbre \cite{peeters2011timbre}. Spectral bandwidth measures the spread of energy around the centroid, complementing it by capturing the noisiness of the spectrum and distinguishing pure tones from broadband content \cite{klapuri2010}.

Feature labels are obtained from sample-library metadata where available. For drums and strings, tempo is taken from the library metadata. For vocals, no such metadata is available so tempo is computed over the whole track using \texttt{librosa.beat\_track}\cite{ellis2007beat, mcfee2015librosa}. For strings and vocals, files are chunked into four second parts and pitch is estimated over the full four second clip using \texttt{librosa.pyin}\cite{mauch2014pyin}. A 100-ms segment most likely to correspond to a stable note is identified as that exhibiting the highest confidence and lowest variation. Its median pitch is stored in the pitch dataset along with the 100-ms segment.

Spectral centroid and spectral bandwidth are also calculated using librosa \cite{mcfee2015librosa} for the full four seconds for the whole 1000 audio files (BPM and pitch balanced) and averaged into one label. A balanced subset of 500 is then used for consistency to avoid artefacts of dataset size. 

Distributions are shown in Table~\ref{tab:dataset_features}. Although well-matched throughout, feature distributions can vary across datasets due to differences in source material. This is a constraint of working with natural datasets that is considered acceptable to work with complex, naturally occurring audio. 

\begin{figure*}
    \centering
    \includegraphics[width=0.95\linewidth]{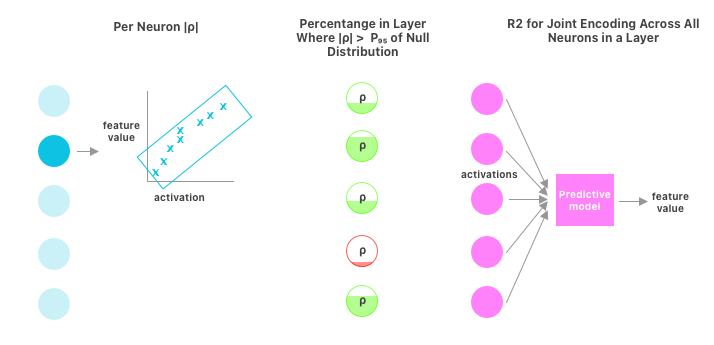}
    \caption{Diagram representing the three metrics used for comparing the relationship of acoustic feature values and neuron activations. $|\rho|$ can be further averaged across layers or models. Null $p_{95}$ derived from permutation tests. Predictive model for $R^2$ can be linear or MLP.}
    \label{fig:metrics}
\end{figure*}

\subsection{Probes} 
\label{sec:methods_probes}  

For each (model, dataset, feature) cell we used three measures to examine relationships between acoustic features and decoder activations (see Fig~\ref{fig:metrics}). Each is computed on 500 audio files balanced for that particular feature (as described above). For each audio segment, we encoded and decoded audio with RAVE, capturing activations at each convolutional unit via forward hooks. 

The first part of our approach follows the linear probing tradition established in \cite{alain2016understanding}, adapted to the generative audio domain. We use Spearman rank correlations between layer activations and acoustic features as a more constrained but more interpretable measure that avoids the capacity concerns raised in probing classifier literature \cite{hewitt2019structural}. This correlational approach has been applied to layer-wise analysis of speech SSL models \cite{chiu2026largescaleprobinganalysisspeakerspecific, pasad2024self}, where it has successfully revealed hierarchical feature encoding. 

For layer level joint encoding, two probe types are used. A linear regression baseline and a nonlinear MLP probe. Reporting both allows us to characterize whether features are encoded in a directly linear form or require nonlinear extraction.

For linear baseline we use a ridge regression with regularization parameter $\alpha = 1.0$. Selected as a baseline because it represents the simplest probe that recovers a linear projection from activations to the target feature. We use a two layer MLP with ReLU activation, followed by a linear output head. The hidden dimension scales with the number of input neurons as \[ h = \min\!\bigl(64,\; \max(8,\; n_{\text{neurons}} / 8)\bigr). \] 

This scaling provides sufficient capacity for layers and clusters with many neurons while preventing severe overparameterization for small neuron subsets. Training uses Adam optimizer with learning rate $10^{-3}$, batch size 64, for 100 epochs with early stopping on validation loss (patience 10). 

Probe performance is evaluated with 5-fold cross-validation. Splits are random across the balanced sample. Predictions are pooled across folds to compute the coefficient of determination ($R^2$) on held-out predictions. The same fold structure is used for linear and nonlinear probes to enable direct comparison. Target features are log-transformed for pitch and spectral centroid (which span multiple octaves and are perceptually logarithmic), and used directly for tempo and spectral bandwidth. $R^2$ values reflect variance explained on the transformed target.

\subsection{Metrics} 
\label{sec:methods_stats}
We characterize feature encoding using three complementary measures (see Fig~\ref{fig:metrics}). Each captures a distinct aspect of how feature information is represented across neurons.

\subsubsection{Encoding}
We firstly provide the per-neuron mean Spearman $|\rho|$ for each neuron. This is the absolute Spearman rank correlation between activation and the target feature, averaged across neurons where appropriate. It provides the average per-neuron encoding strength.

Prevalence above null (\% $>p_{95}$) is the percentage of neurons whose $|\rho|$ exceeds the 95th percentile of a permutation-derived null distribution. This measures how broadly distributed feature-encoding neurons are within the population. 

We finally provide a joint probe $R^2$ to assess the performance of a multi-neuron probe (linear or nonlinear) on the neuron vector as a whole. This measures the joint-encoding capacity of the population by telling us how well the feature is recoverable from the full set of neurons together.

For each (model, dataset, feature) cell, we construct an empirical null distribution by permuting the target feature labels across samples and recomputing all three measures. Permutations are repeated $K=500$ for Spearman correlations and $K=100$ times for nonlinear probes (which are themselves each 5-fold cross-validated). For each measure, we record the mean and 95th percentile of the null distribution. The 95th percentile defines the chance threshold against which the observed values are compared.

\subsubsection{Paired Comparisons}
For paired comparisons across cells (natural vs synthetic), we use Wilcoxon signed-rank tests where pairing is matched by layer position and model.

Pairing within layer is deliberate because encoding strength varies strongly and non-monotonically with depth, comparing natural vs synthetic values at the same layer removes layer as a confound and isolates the effect of distributional match from the effect of network depth. Across the three models, this yields 84 layer-matched pairs per feature, which we tested with a two-sided Wilcoxon signed-rank test and report matched-pairs rank-biserial correlation $r$ as the effect size. We note that pairs sharing a model are not strictly independent, as they inherit a common model-level component. We report the between-model intraclass correlations (ICC) and the exact sign-permutation $p$ stratified by each factor.

\subsubsection{Depth}
Where we test whether layer position predicts encoding strength, for each (model, dataset) cell, we fit polynomial regressions of the metric on normalized layer depth $d \in [0, 1]$, including both linear ($\beta_0 + \beta_1 d$) and quadratic ($\beta_0 + \beta_1 d + \beta_2 d^2$) specifications. Slopes from each cell are treated as the unit of observation with the curvature coefficient ($\beta_2$) tested across cells with a one-sample Wilcoxon signed-rank test. Per-cell regressions are well-conditioned (each cell contributes 28 layer observations).

We report the matched-pairs rank-biserial correlation $r$ as the effect size. Cells are not fully independent as they share both models and datasets so we report the between-model and between-dataset ICC and the exact sign-permutation $p$ stratified by each factor.

\subsubsection{Inferential and Descriptive}
The confirmatory analyses specified by our research questions are the natural-vs-synthetic probe comparison (RQ1), the depth-dependence of encoding for natural audio (RQ2), and the cluster-versus-layer comparison (RQ3). Exploratory analyses are reported descriptively without inferential claims and include synthetic-stimulus depth profiles (3 cells, insufficient for inference). The comparison of models (RAVE and EnCodec) is also non-inferential as the difference in structure makes per-layer matching not possible and low N makes cell-level comparisons equally weak.

We control the false-discovery rate (Benjamini--Hochberg) within each family of related tests (a given measure across the four features) and report adjusted p-values. The three measures are treated as complementary views of the same representation rather than substitutable tests of a single hypothesis. Confidence intervals are 95\% bootstrap.

\subsection{Cross-Layer Clustering} 
To identify functionally-coherent neuron groups beyond individual layers, we performed hierarchical cross-layer clustering. We divided the 21 decoder layers into three sections based on architectural depth: early (layers 0-7), middle (layers 8-14), and late (layers 15+). This sectioning reflects both architectural structure (early layers process compressed latents, late layers generate waveforms) and our per-layer findings (different metrics peak at different depths).

Within each section, we used K-means clustering on the selected neurons based on their activation patterns across the audio corpus. Each neuron's feature vector consists of its activations across the full set of stimuli independently of any probed feature. We used $k = 6$ clusters to group neurons with similar response profiles (see Supplementary Materials Fig~S2 to~S4 for more details on the choice of $k$).

We apply the same layer level metrics described above by recalculating the observed mean $|\rho|$ and \% $>p_{95}$ and retraining new linear and nonlinear probes to the new neuron subset to investigate the cluster's joint encoding. 

For each cell, feature, and measure, we identified the best-performing cross-layer cluster and the best-performing single layer within that section, and computed their difference. Differences were tested across cells using the same framework as the depth analysis where the cell is the unit of observation, and ICC for model and dataset is reported. Because clusters and their best-matching layer are selected from overlapping neuron pools, the two sets may not be disjoint. However, this does not undermine the paired comparison as shared neurons contribute equally to both members of a pair and so cancel during within-cell differencing. 


\subsection{EnCodec Comparison}

We also ran the above evaluations on Meta's EnCodec model \cite{defossez2022highfidelity}. We chose the pretrained 32 kHz model designed for music and ran all four datasets through, collecting activations and calculating each of the three metrics across the model's layers. It is structurally similar (input convolution, upsampling stages each followed by a residual block, output convolution), however, it uses residual vector quantization rather than variational continuous latents. We identified 13 appropriate layers to capture activations. This comparison allows us to see if results hold across a different decoder architecture.
\section{Results}

\begin{table*}[t]
\centering
\caption{Encoding metrics aggregated by feature and condition, reported mean across cells. Synthetic spectral values not analyzed. $\times$ null is ratio of observed mean against permutation derived null mean.} 
\label{tab:encoding_summary}
\renewcommand{\arraystretch}{1.15}
\setlength{\tabcolsep}{5pt}
\small
\begin{tabular}{l l r | r r r | r r r r}
\toprule
& & & \multicolumn{3}{c|}{\textbf{Per-neuron}} 
  & \multicolumn{4}{c}{\textbf{Joint probe $R^2$}} \\
\textbf{Feature} & \textbf{Condition} & $n$
  & Mean $|\rho|$ & $\times$ null & \% $>$ p95
  & Linear & Nonlinear & $\times$ null & NL gain \\
\midrule
\multirow{4}{*}{Pitch}
  & Natural   & 6 & 0.20 & 2.24 & 56.4 & 0.27 & 0.38 & 13.18 & 0.11 \\
 & Synthetic & 3 & 0.45 & 5.12 & 82.3 & 0.99 & 1.00 & 35.31 & 0.00 \\
 & ID        & 2 & 0.18 & 2.10 & 50.5 & 0.17 & 0.43 & 14.41 & 0.11 \\
 & OOD       & 4 & 0.20 & 2.32 & 59.3 & 0.25 & 0.36 & 12.55 & 0.11 \\
\midrule
\multirow{4}{*}{BPM}
  & Natural   & 9 & 0.20 & 2.29 & 59.7 & 0.27 & 0.42 & 13.56 & 0.15 \\
 & Synthetic & 3 & 0.76 & 8.63 & 91.6 & 1.00 & 0.99 & 28.57 & 0.00 \\
 & ID        & 3 & 0.19 & 2.17 & 55.4 & 0.32 & 0.54 & 17.11 & 0.22 \\
 & OOD       & 6 & 0.20 & 2.35 & 61.9 & 0.25 & 0.36 & 11.76 & 0.12 \\
\midrule
\multirow{3}{*}{Spectral Centroid}
  & Natural & 9 & 0.31 & 3.45 & 69.8 & 0.61 & 0.76 & 22.48 & 0.15 \\
 & ID      & 3 & 0.26 & 2.92 & 61.5 & 0.63 & 0.80 & 24.69 & 0.17 \\
 & OOD     & 6 & 0.33 & 3.72 & 73.9 & 0.61 & 0.74 & 21.44 & 0.13 \\
\midrule
\multirow{3}{*}{Spectral Bandwidth}
  & Natural & 9 & 0.27 & 3.15 & 69.9 & 0.47 & 0.68 & 21.52 & 0.20 \\
 & ID      & 3 & 0.24 & 2.73 & 65.9 & 0.50 & 0.75 & 25.60 & 0.25 \\
 & OOD     & 6 & 0.29 & 3.36 & 71.9 & 0.46 & 0.64 & 19.68 & 0.18 \\
\bottomrule
\end{tabular}
\end{table*}

\subsection{Synthetic Stimuli}

\begin{figure*}
    \centering
    \includegraphics[width=0.91\linewidth]{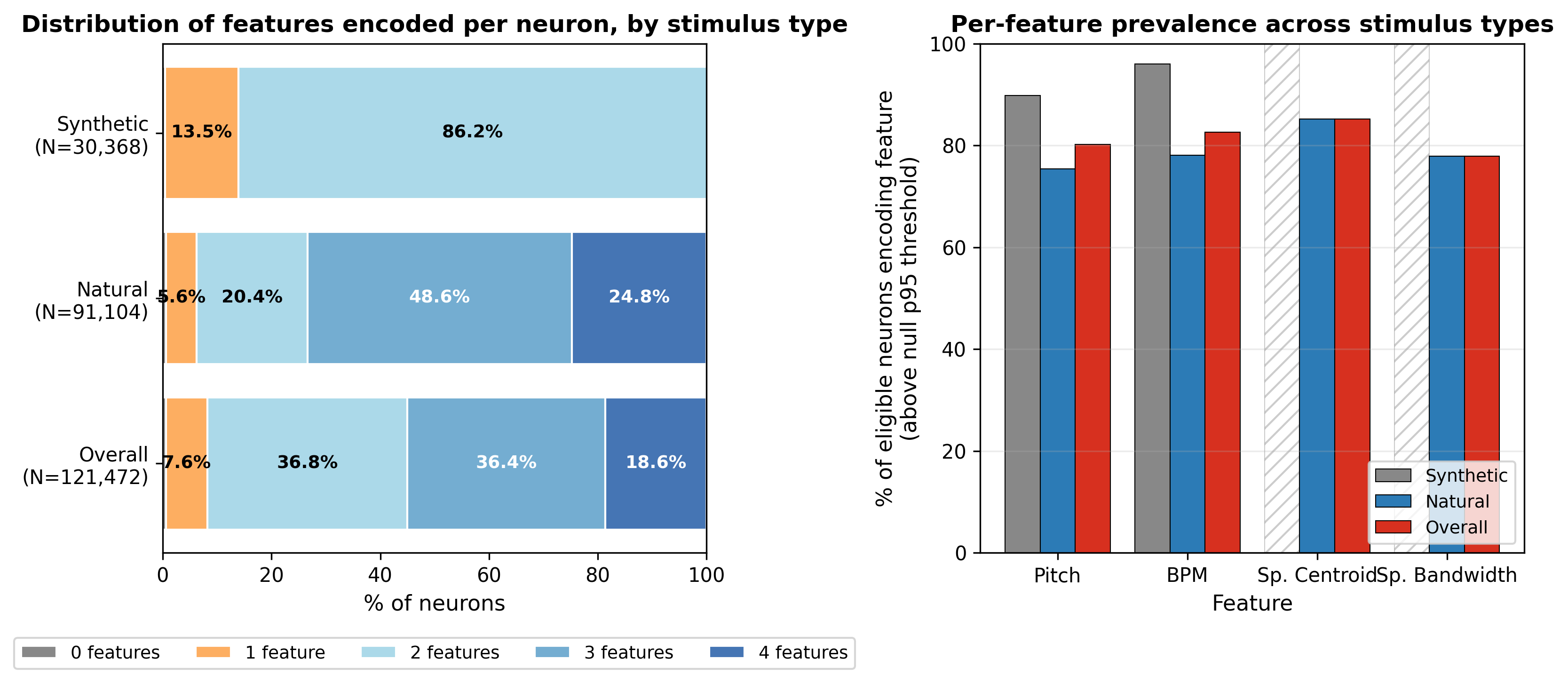}
    \caption{Plots demonstrating what percentage of neurons encode multiple features across all cells. Left shows how many features are $>p_{95}$ for each neuron; synthetic can have max.\ 2 features as this is all that is measured. Right shows what percentage of eligible neurons are $>p_{95}$ for each feature (N=121,472 for BPM (all data), N=91,104 for pitch (no drums), N=91,104 for spectral centroid and spectral bandwidth (no synthetic)).}
    \label{fig:specialisation}
\end{figure*}

Shown in Table~\ref{tab:encoding_summary}, we found that for pitch and BPM (the two features measured), synthetic stimuli showed high $|\rho|$ (0.45 and 0.76) showing a correlation substantially above chance between activation and features at a neuron level. Almost all neurons responded above chance (\% $>p_{95}$ is 82\% and 92\%) and further shown in Fig~\ref{fig:specialisation}, 86.2\% of neurons responded to both and less than 1\% neither. We see the encoding is mostly captured linearly and that joint encoding for $R^2$ is near the ceiling. This likely reflects the stimuli's controlled single-dimension variation, which removes the confounding co-variation present in natural audio and makes each feature easily recoverable.

\subsection{Natural Audio}

For natural audio, we see again in Table~\ref{tab:encoding_summary} that all features are correlated above chance with $|\rho|$ between 0.20 and 0.31. \% $>p_{95}$ is also high. We also see a much larger gain when analysing the joint encoding between linear and nonlinear models than we saw with synthetic inputs. $R^2$ is well above chance. Fig~\ref{fig:specialisation} shows encoding is often shared with 73.4\% of neurons correlating over chance for at least 3 features.

When directly comparing synthetic and natural audio for $|\rho|$ and \% $>p_{95}$ using layer matched paired Wilcoxon signed rank tests, synthetic stimuli were encoded far more strongly than natural audio across all measures for pitch (100\% of layer-matched pairs saw rank-biserial r $\geq$ 0.999; all $p<0.001$). Between-model ICC was moderate to high (0.097--0.835), reflecting variation in the magnitude of the synthetic advantage across models rather than any disagreement in direction (the effect direction was unanimous across all models). 

\subsection{Out-of-Distribution Audio}
Feature encoding remains well above chance when a model was tested with out-of-distribution audio it had not been trained on. Out-of-distribution per-neuron correlations exceeded the null threshold by 2--4$\times$ and joint $R^2$ remained high (e.g.\ spectral centroid $R^2 = 0.74$, tempo $R^2$ = 0.36 on unfamiliar audio), indicating the learned feature representations generalize substantially beyond the training domain. Notably, on per-neuron measures encoding was often as strong or stronger out-of-distribution than in-distribution, suggesting the network's individual channels correlate with these features in a manner not specific to the training data.

\subsection{Effect of Layer Depth}

\begin{table*}[t]
\centering
\caption{Median per-cell quadratic coefficient $\beta_2$ over normalized depth (positive $=$ U-shaped, negative $=$ inverted-U) with [IQR]. $r$ is the matched-pairs rank-biserial effect size. Between-model and between-dataset intraclass correlations (ICC) are provided, along with the Benjamini--Hochberg-adjusted exact sign-permutation $p$. All natural cells; synthetic omitted (descriptive, $N{=}3$).}
\label{tab:depth_main}
\renewcommand{\arraystretch}{1.2}\setlength{\tabcolsep}{4pt}\footnotesize
\begin{tabular}{l l r r r r r r r c}
\toprule
& & & & & \multicolumn{2}{c}{\textbf{ICC}} & \\
\cmidrule(lr){6-7}
\textbf{Meas.} & \textbf{Feature} & $N$ & $\beta_2$ [IQR] & $r$ & model & data & $p_{\text{adj}}$  \\
\midrule
\multirow{4}{*}{Mean $|\rho|$}
  & Pitch               & 6 & 0.11 [0.09, 0.17]  & 1.00  & 0.37 & 0.00 & 0.042* \\
  & BPM                 & 9 & -0.09 [-0.11, 0.03] & -0.38 & 0.00 & 0.00 & 0.324 \\
  & Spectral Centroid   & 9 & 0.25 [0.16, 0.55]  & 0.78  & 0.74 & 0.00 & 0.042* \\
  & Spectral Bandwidth  & 9 & 0.43 [0.17, 0.57]  & 0.87  & 0.32 & 0.00 & 0.042* \\
\midrule
\multirow{4}{*}{\% $>$ p95}
  & Pitch               & 6 & 26.16 [12.34, 43.61]  & 0.81  & 0.06 & 0.01 & 0.125 \\
  & BPM                 & 9 & -12.79 [-24.61, 8.96] & -0.38 & 0.00 & 0.00 & 0.250 \\
  & Spectral Centroid   & 9 & 16.42 [1.41, 41.15]   & 0.64  & 0.67 & 0.00 & 0.125 \\
  & Spectral Bandwidth  & 9 & 41.62 [24.16, 47.15]  & 0.96  & 0.00 & 0.00 & 0.031* \\
\midrule
\multirow{4}{*}{$R^2$}
  & Pitch               & 6 & -0.38 [-0.45, -0.13] & -1.00 & 0.00 & 0.72 & 0.031* \\
  & BPM                 & 9 & -0.65 [-1.01, -0.55]  & -1.00 & 0.00 & 0.00 & 0.005** \\
  & Spectral Centroid   & 9 & -0.29 [-0.34, -0.20]  & -1.00 & 0.70 & 0.00 & 0.005** \\
  & Spectral Bandwidth  & 9 & -0.38 [-0.60, -0.35]  & -1.00 & 0.43 & 0.00 & 0.005** \\
\bottomrule
\end{tabular}
\end{table*}

\begin{figure*}
    \centering
    \includegraphics[width=0.95\linewidth]{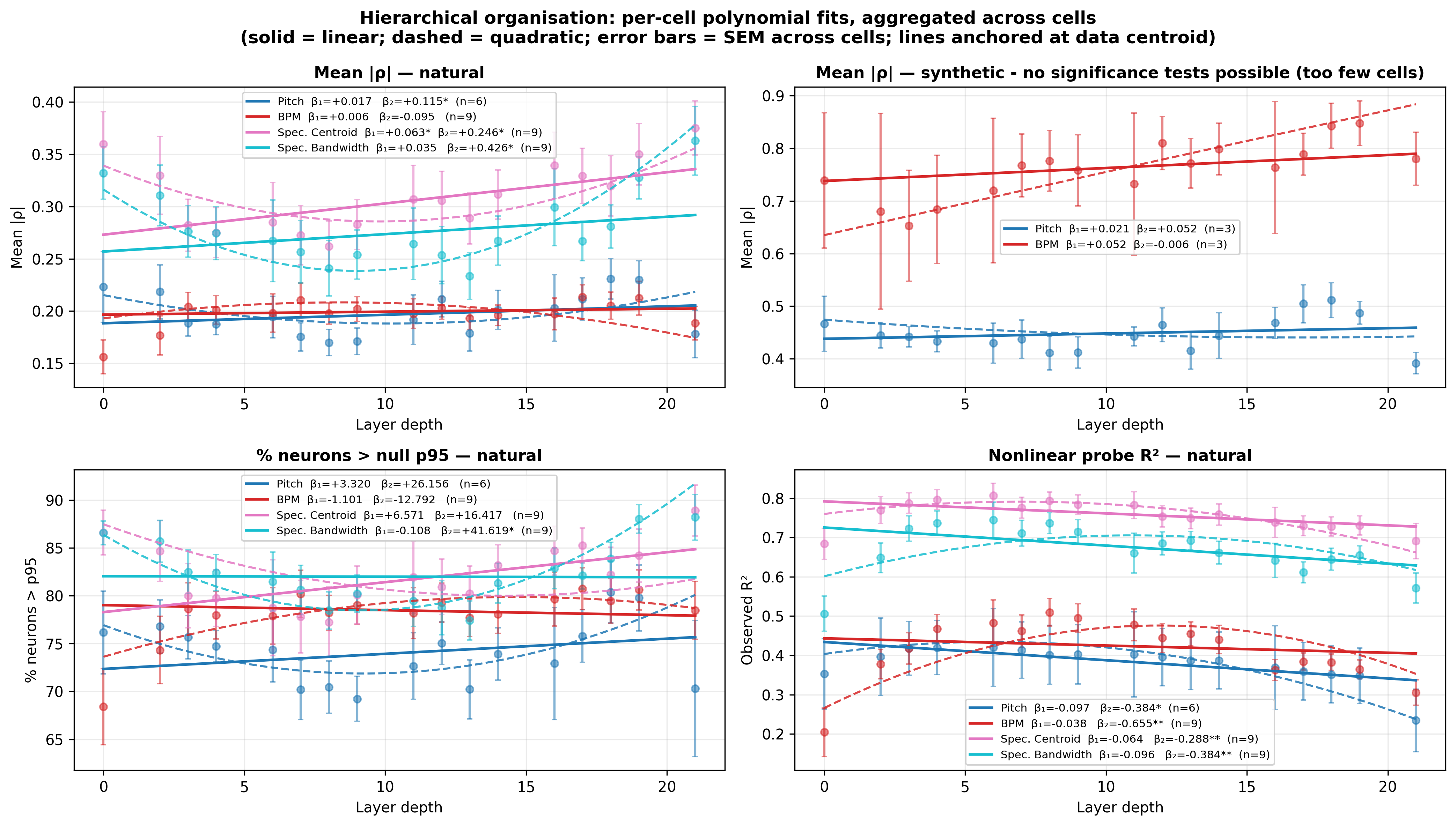}
    \caption{Plots of per layer mean $|\rho|$, neuron prevalence and $R^2$ across four audio features for natural audio, reported as median across cells. Only per layer mean $|\rho|$ displayed for synthetic audio as other features are near ceiling throughout (e.g.\ flat). Significance values come from testing curvature coefficient ($\beta_2$) across cells with a one-sample Wilcoxon signed-rank test. Not possible for synthetic as N=3 (too low).}
    \label{fig:depth}
\end{figure*}

Depth analysis is conducted on natural audio cells, where the per-cell aggregation has adequate statistical power. For synthetic stimuli, joint encoding is at ceiling across all layers ($R^2 > 0.99$), and depth-related variation is too small to be meaningful given the small sample of synthetic cells (N=3 per feature). Results are shown in Fig~\ref{fig:depth} and synthetic $|\rho|$ is still included to show its similar pattern to natural audio, although we are unable to statistically determine significance.

Joint $R^2$ followed an inverted-U depth profile (negative curvature, peak in mid-network) for every feature: pitch ($-0.38$; $N=6$, $p=0.031$, $r=-1.00$), BPM (median $\beta_2 = -0.65$, $N=9$, $p=0.005$, $r=-1.00$), spectral centroid ($-0.29$; $N=9$, $p=0.005$, $r=-1.00$) and spectral bandwidth ($-0.38$; $N=9$, $p=0.005$, $r=-1.00$). 

The effect was unanimous in direction across models and datasets. BPM showed no between-group dependence (both ICCs =0), while pitch, spectral centroid, and spectral bandwidth carried moderate ICC on one grouping axis each (Table~\ref{tab:depth_main}), reflecting variation in the magnitude of the curvature across groups rather than any disagreement in its direction. The inverted-U is therefore robust as a directional finding but its precise magnitude is less certain for the high-ICC features.

Per-neuron mean $|\rho|$ followed the opposite, U-shaped profile (positive curvature, strongest at network extremes) for pitch (median $\beta_2 = +0.11$; $N=6$, $p=0.042$, $r=+1.00$), spectral centroid ($+0.25$; $N=9$, $p=0.042$, $r=+0.78$), and spectral bandwidth ($+0.43$; $N=9$, $p=0.042$, $r=+0.87$); tempo showed no significant curvature. ICC shows moderate effect of model but 0 for dataset.

We examined the five top layers in each cell to determine if there were any layers that consistently encoded particular features across combinations, however, no specific layers could be determined as shared when corrected for the multiple tests.

\subsection{Cross-Layer Clustering}

For comparing clusters, we pick the best cluster per section (after filtering out clusters that are less than 1\% of the total available) and compare this against the best layer in the corresponding section. Filtering prevents best selection picking a small cluster (approx.\ 1--5 neurons) that may have high correlation by chance.

\begin{figure*}
    \centering
    \includegraphics[width=1.0\linewidth]{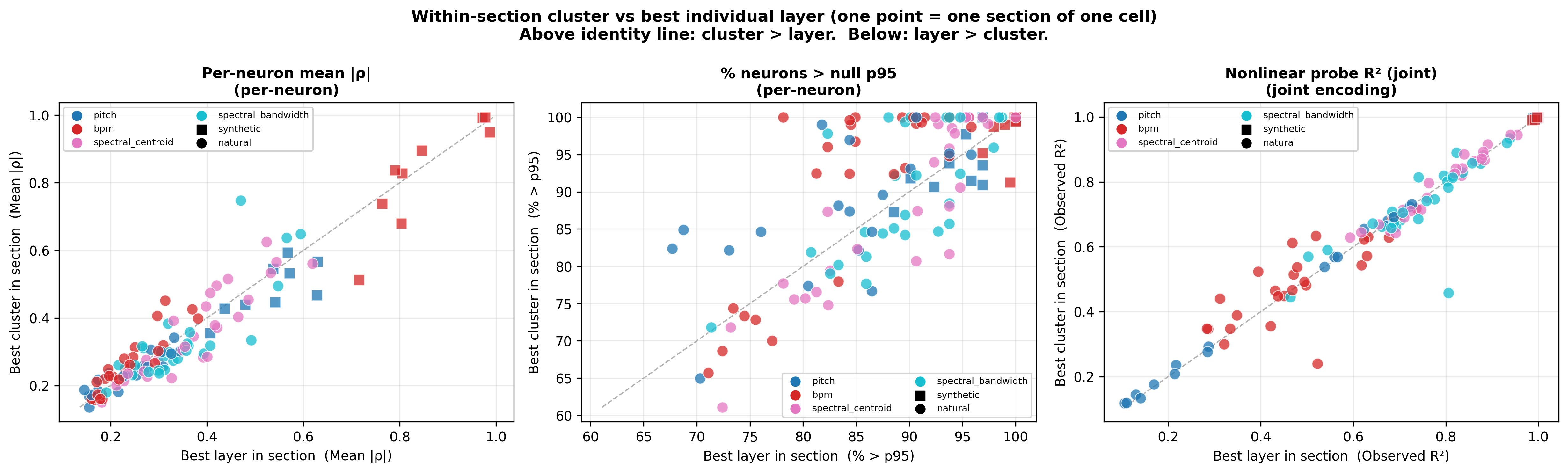}
    \caption{Best cluster against best layer per section for each cell for all three metrics ($|\rho|$, \% $>p_{95}$ and $R^2$); only natural BPM $|\rho|$ and $>p_{95}$ show improvement for clusters.}
    \label{fig:cluster_scatter}
\end{figure*}

We then compare the deltas across each cell (full results in Table~S4 see also Fig~\ref{fig:cluster_scatter}). Comparing each cell's best cross-layer cluster against its best single layer, tempo (BPM) was the only feature to show a reliable cluster advantage, and it did so across two measures: per-neuron mean~$|\rho|$ (HL${}=+0.019$, $r=+0.65$), \% $>p_{95}$ ($+6.15$, $r=+0.75$). Both showed low to moderate dataset ICC but no effect of model. High concordant $r$ indicates a consistent direction across cells but a magnitude that varies by dataset.

\subsection{EnCodec Comparison}

Table~\ref{tab:encoding_summary_encodec} demonstrates that findings from RAVE models are well replicated in another architecture. Synthetic stimuli show strong linear encoding, with BPM encoding with more strength and more prevalence than pitch. Synthetic stimuli were encoded more strongly than natural audio across every measure and feature, replicating the pattern observed in RAVE. For pitch: mean $|\rho|$ ($0.56$ vs $0.24$), \% $>p_{95}$ ($85.7\%$ vs $64.6\%$) and joint $R^2$ ($0.99$ vs $0.45$). For tempo: mean $|\rho|$ ($0.66$ vs $0.20$), \% $>p_{95}$ ($94.0\%$ vs $55.6\%$) and joint $R^2$ ($0.99$ vs $0.59$). All comparisons $N=13$, $p<0.001$, $r=+1.00$.

As with RAVE, spectral centroid had the highest $|\rho|$, with good nonlinear joint encoding (nonlinear gain strong for BPM ($+0.20$), present but weaker for other features). Given the low number of cells we are only able to provide descriptive analysis of the effect of depth, but we do see in Table~S3 that $\beta_2$ is all negative for $R^2$, indicating that joint encoding peaks at intermediate decoder layers. We also see similar peaks in later layers for $|\rho|$ (spectral centroid, spectral bandwidth, BPM for natural, pitch for synthetic). We also see little improvement in joint encoding for clustering.

\begin{table*}[t]
\centering
\caption{EnCodec encoding metrics aggregated by feature and condition, results are mean across cells. Multiples ($\times$ null) are the observed value divided by the corresponding null baseline ($p95$ for per-neuron, mean null $R^2$ for the joint probe).}
\label{tab:encoding_summary_encodec}
\renewcommand{\arraystretch}{1.15}
\setlength{\tabcolsep}{5pt}
\small
\begin{tabular}{l l r | r r r | r r r r}
\toprule
& & & \multicolumn{3}{c|}{\textbf{Per-neuron}} 
  & \multicolumn{4}{c}{\textbf{Joint probe $R^2$}} \\
\textbf{Feature} & \textbf{Condition} & $n$
  & Mean $|\rho|$ & $\times$ null & \% $>$ p95
  & Linear & Nonlinear & $\times$ null & NL gain \\
\midrule
\multirow{2}{*}{Pitch}
  & Natural   & 2 & 0.24 & 2.76 & 64.6 & 0.44 & 0.45 & 18.03 & 0.05 \\
& Synthetic & 1 & 0.56 & 6.41 & 85.7 & 0.99 & 0.99 & 35.91 & 0.01 \\
\midrule
\multirow{2}{*}{BPM}
  & Natural   & 3 & 0.20 & 2.26 & 55.6 & 0.46 & 0.59 & 21.85 & 0.20 \\
& Synthetic & 1 & 0.66 & 7.45 & 94.0 & 1.00 & 0.99 & 32.81 & 0.00 \\
\midrule
\multirow{1}{*}{Spectral Centroid}
  & Natural & 3 & 0.33 & 3.71 & 65.9 & 0.83 & 0.85 & 33.80 & 0.07 \\
\midrule
\multirow{1}{*}{Spectral Bandwidth}
  & Natural & 3 & 0.26 & 2.98 & 63.6 & 0.79 & 0.82 & 26.14 & 0.09 \\
\bottomrule
\end{tabular}
\end{table*}

\section{Discussion}
\subsection{Robustness to Complexity}
When a synthetic stimulus varies along a single axis (pitch or BPM, with all else fixed), the corresponding RAVE activations vary in a linear manner with that axis. This is a strong result and it implies that RAVE has learned a representation where the feature has a near-linear coordinate in activation space and that small input changes produce proportional activation changes.

Beyond this, the substantial encoding of natural audio despite its acoustic complexity demonstrates that RAVE's decoder also maintains detectable feature correlations under realistic conditions. While synthetic stimuli produced significantly stronger encoding, natural audio still showed meaningful correlations, indicating the network's activations track features even in signals containing timbral variation, temporal irregularities, and polyphonic content.

When considering differing responses, we could also characterize synthetic and natural stimuli as triggering different aspects of the same representation. Synthetic stimuli demonstrate what the network's representation is like if all variation aligned with a single feature axis. Natural stimuli show what the representation is when feature variation is entangled with other natural sources of variation. This is perhaps why it requires nonlinear probing to uncover. 

\subsection{Feature Specialization}

Neuron-level encoding is multi-feature. In natural audio, 58.7\% of neurons encode both pitch and tempo above the null threshold, and 73\% encode at least three of the four analyzed features. This overlap indicates that RAVE encodes acoustic features in a distributed, entangled manner rather than allocating dedicated neurons to individual properties. A single neuron typically contributes to the representation of several features at once.

This being said, spectral centroid shows the highest per-neuron correlation and encoding across natural cells, suggesting that brightness is the most neuron-aligned acoustic property in the network's representation.

The synthetic-to-natural comparison reveals asymmetry between pitch and tempo. Under controlled synthetic stimuli, tempo was the more strongly encoded of the two, yet on natural audio the two converged to the same value. The degradation of tempo from synthetic to natural conditions is far steeper than for pitch. 

This could be a reflection of the different ways the two properties exist in the two formats. In natural audio, tempo must be inferred from irregular, expressively-timed onsets embedded in polyphonic textures. This is a far noisier estimation problem than reading a fixed periodicity. 

Pitch, by contrast, is more of a locally-instantiated property. Although it too weakens on natural audio, its encoding is less dependent on the regularity that the synthetic condition supplies. Therefore, whilst it is an interesting result, we can caution against reading synthetic encoding strength as a direct proxy for how a model represents tempo in realistic use.

That tempo alone benefits from cross-layer aggregation is consistent with its status as the only temporally-extended feature examined. Unlike other features, tempo is defined over time and its encoding may draw on layers operating at different timescales, so that grouping neurons across layers recovers information no single layer holds in isolation.

\subsection{Joint-Layer and Per-Neuron Encoding}
In some cases, we see dissociation between metrics, especially $|\rho|$ and $R^2$. These metrics measure different qualities of layers, and so their divergence can highlight differing behaviors, especially when observed across different model depths. 

Indeed, our the depth analysis indicates these different layer level patterns. A plausible explanation is that the decoder distributes feature information across many neurons in its middle layers, where joint recovery ($R^2$) is strongest, and consolidates it into fewer dominant neurons closer to the output, where per-neuron correlations ($|\rho|$) are strongest and joint diversity is reduced.

\subsection{Application to Other Architectures}
RAVE \cite{caillon2021rave} is a deliberate choice for this initial investigation rather than a limiting one. Its decoder is a relatively shallow stack of transposed convolutions and residual blocks operating on a low-dimensional latent making it an architecture simple enough to permit exhaustive per-layer analysis. However, it is also representative of the convolutional VAE family that remains widely deployed in real-time musical applications where latency constraints preclude larger autoregressive or diffusion-based alternatives. The questions of whether musical features occupy identifiable locations in a generative decoder and whether cross-layer functional groupings improve feature readout are architecture-agnostic in formulation, even if our specific answers are not.

We expect the methodological framework (permutation-validated prevalence alongside encoding strength, cross-layer clustering, neuron specialization taxonomy) to transfer directly to other architectures, and the broad findings about how features are organized to plausibly generalize to other convolutional decoders which share RAVE's upsampling-through-strided-convolution backbone. We have shown this is the case for EnCodec \cite{defossez2022highfidelity} and the results with this general purpose decoder align with our strong results for OOD audio in RAVE models.

We are more cautious about generalization to architectures with fundamentally different inductive biases. Transformer-based audio models such as MusicGen \cite{copet2023simple} and AudioLM \cite{borsos2023audiolmlanguagemodelingapproach} replace local convolutional receptive fields with global attention. This would likely redistribute feature encoding across token positions rather than network depth. We see this contrast paralleled in vision research, where layer-wise interpretability findings for convolutional networks do not transfer cleanly to vision transformers \cite{raghu2021vision}. 

Diffusion-based synthesizers (\cite{kong2020diffwave}, \cite{gong2025acestepstepmusicgeneration}) introduce an additional temporal axis (denoising steps) along which features may be organized, complicating direct comparison to a feed-forward decoder hierarchy. Autoregressive token models operate on discrete codebook entries rather than continuous activations, which changes what "neuron-level encoding" even means in those settings \cite{van2017neural}. Specific localization claims should therefore be read as findings about VAE-family decoders, while the methodology and the higher-level organizational principles offer hypotheses to be tested in these other settings rather than conclusions to be assumed of them.

\section{Limitations}

Our correlational analysis establishes association between layer activations and musical features, not functional importance. A neuron that correlates with pitch may actually be tracking a correlated acoustic property. For example, in an extended dissection of the MusiCNN tagger Sturm~\cite{sturm2022musicnn} examines a single network on individual signals across its convolutional stack. Identical musical excerpts drawn from different source recordings yield markedly different tag predictions, and predictions shift with input amplitude alone.  

The drum model is excluded from pitch analysis and although not tonal nor melodic in the traditional sense, drums do contain spectral content (e.g., toms, cymbals). Future work could investigate whether pitch encoding of percussive timbres differs from tonal instruments. While permutation testing confirms results are well above chance (approx. 2--4$\times$ the null mean across features), the effect sizes observed for natural audio encoding are modest (pitch mean $|\rho|$=0.20, tempo mean $|\rho|$=0.20, spectral centroid mean $|\rho|$=0.31, spectral bandwidth mean $|\rho|$=0.27), corresponding to approximately 4--10\% explained variance. 

\section{Conclusion}

We have presented a systematic layer-wise and cross-layer cluster analysis of spectro-temporal feature encoding in RAVE audio synthesis models, examining three models trained on different musical domains across four stimulus types using three complementary metrics: encoding strength, permutation-validated feature prevalence and joint-encoding strength with a nonlinear probe. We also conducted these tests on a supplementary model to test the generalization of results beyond one specific architecture.

We have found that synthetic stimuli are strongly encoded throughout the decoder layers and these results hold for natural audio and out-of-distribution audio and across temporal, pitched and spectral features. Synthetic stimuli are strongly linearly correlated whereas natural audio shows stronger results when measuring joint encoding across a whole layer with a nonlinear probe. Monotonic encoding ($|\rho|$) is strongest at the early and late layers for three out of four features, whereas joint encoding is strongest in the middle layers for all features. Clustering across layers shows small but consistent advantage when finding neuron subsets that encode tempo, in comparison to whole layers.

Future work should pursue several directions: (1) causal validation through targeted ablation of the identified clusters and layers, which would determine whether the correlational patterns reported here reflect functional roles; (2) further cross-architecture comparisons to determine whether these findings generalize beyond VAEs; (3) investigation of whether the identified encoding structure can inform practical applications such as targeted network bending, structured pruning, or the design of more interpretable neural audio architectures.

\section*{Acknowledgments}
The authors would like to thank the developers and maintainers of the RAVE library.

\section*{Supplementary Materials}
This paper has supplementary downloadable material available at https://ieeexplore.ieee.org/Xplore provided by the author. The material includes  additional plots (Figs S1–S4) and data tables (Tables S1–S4). Contact l.mccallum@arts.ac.uk for further questions about this work.

\bibliographystyle{IEEEtran}
\bibliography{library}

\end{document}


\onecolumn
\clearpage
\section*{Supplementary Materials}
\subsection{Permutation Tests}
\label{app_permutation}

\begin{table}[!h]
\centering
\caption{Per-cell encoding metrics for all (model, dataset, feature) combinations.
Per-neuron statistics from Spearman correlations; joint statistics from MLP probes.
Null \% $>p_{95}$ values are from 500 permutations of shuffled feature labels.
Linear $R^2$ from ridge regression probes; nonlinear $R^2$ from MLP probes;
NL gain = nonlinear $-$ linear $R^2$. Bracketed values are 95\% bootstrap confidence intervals calculated at a layer level.}
\label{tab:supp_per_cell}
\renewcommand{\arraystretch}{1.10}
\setlength{\tabcolsep}{4pt}
\scriptsize
\begin{tabular}{l l l | r r r r | r r r r}
\toprule
& & & \multicolumn{4}{c|}{\textbf{Per-neuron $|\rho|$}}
  & \multicolumn{4}{c}{\textbf{Joint probe $R^2$}} \\
\textbf{Model} & \textbf{Dataset} & \textbf{Feature}
  & Null $p_{95}$ & Obs. mean [95\% CI] & Obs. max & \% $>$ p95
  & Null $p_{95}$ & Linear [95\% CI] & Nonlinear [95\% CI] & NL gain \\
\midrule
\multirow{13}{*}{Strings}
  & \multirow{4}{*}{Strings} & Pitch & 0.087 & 0.221 [0.210, 0.233] & 0.729 & 58.6 & 0.023 & 0.460 [0.436, 0.475] & 0.619 [0.584, 0.635] & 0.159 \\
  & & BPM & 0.087 & 0.153 [0.148, 0.158] & 0.548 & 45.1 & 0.037 & 0.377 [0.323, 0.418] & 0.512 [0.480, 0.541] & 0.134 \\
  & & Spec C. & 0.088 & 0.367 [0.349, 0.383] & 0.904 & 77.4 & 0.038 & 0.817 [0.787, 0.838] & 0.908 [0.893, 0.921] & 0.091 \\
  & & Spec B. & 0.088 & 0.256 [0.239, 0.277] & 0.838 & 65.2 & 0.026 & 0.751 [0.706, 0.786] & 0.836 [0.784, 0.866] & 0.086 \\
\cmidrule{2-11}
  & \multirow{3}{*}{Drums} & BPM & 0.085 & 0.172 [0.166, 0.176] & 0.431 & 64.5 & 0.027 & 0.246 [0.201, 0.277] & 0.329 [0.286, 0.361] & 0.083 \\
  & & Spec C. & 0.087 & 0.376 [0.366, 0.384] & 0.776 & 87.1 & 0.034 & 0.410 [0.349, 0.435] & 0.641 [0.624, 0.654] & 0.230 \\
  & & Spec B. & 0.086 & 0.517 [0.501, 0.531] & 0.857 & 92.8 & 0.039 & 0.503 [0.479, 0.525] & 0.681 [0.662, 0.696] & 0.178 \\
\cmidrule{2-11}
  & \multirow{2}{*}{Stimuli} & Pitch & 0.088 & 0.468 [0.442, 0.500] & 0.980 & 81.8 & 0.027 & 0.993 [0.991, 0.995] & 0.994 [0.992, 0.995] & 0.001 \\
  & & BPM & 0.092 & 0.945 [0.932, 0.953] & 0.998 & 99.3 & 0.041 & 0.998 [0.997, 0.998] & 0.994 [0.992, 0.994] & $-$0.004 \\
\cmidrule{2-11}
  & \multirow{4}{*}{Vocals} & Pitch & 0.088 & 0.134 [0.128, 0.141] & 0.440 & 42.9 & 0.035 & 0.066 [0.058, 0.074] & 0.125 [0.113, 0.140] & 0.059 \\
  & & BPM & 0.089 & 0.177 [0.170, 0.185] & 0.468 & 55.3 & 0.035 & 0.202 [0.171, 0.227] & 0.341 [0.318, 0.361] & 0.139 \\
  & & Spec C. & 0.088 & 0.315 [0.308, 0.323] & 0.774 & 83.4 & 0.040 & 0.454 [0.417, 0.480] & 0.614 [0.590, 0.642] & 0.160 \\
  & & Spec B. & 0.087 & 0.190 [0.176, 0.206] & 0.725 & 52.6 & 0.030 & 0.402 [0.375, 0.418] & 0.570 [0.530, 0.615] & 0.169 \\
\midrule
\multirow{13}{*}{Drums}
  & \multirow{4}{*}{Strings} & Pitch & 0.087 & 0.264 [0.252, 0.276] & 0.720 & 72.7 & 0.025 & 0.354 [0.333, 0.365] & 0.522 [0.503, 0.533] & 0.168 \\
  & & BPM & 0.087 & 0.219 [0.194, 0.239] & 0.629 & 50.5 & 0.039 & 0.440 [0.422, 0.455] & 0.447 [0.425, 0.463] & 0.008 \\
  & & Spec C. & 0.088 & 0.464 [0.448, 0.486] & 0.897 & 87.3 & 0.030 & 0.803 [0.791, 0.811] & 0.847 [0.834, 0.858] & 0.044 \\
  & & Spec B. & 0.089 & 0.245 [0.233, 0.259] & 0.751 & 74.8 & 0.040 & 0.570 [0.540, 0.594] & 0.711 [0.681, 0.740] & 0.142 \\
\cmidrule{2-11}
  & \multirow{3}{*}{Drums} & BPM & 0.088 & 0.187 [0.174, 0.197] & 0.577 & 55.1 & 0.030 & 0.165 [0.112, 0.225] & 0.536 [0.452, 0.580] & 0.370 \\
  & & Spec C. & 0.090 & 0.236 [0.221, 0.256] & 0.816 & 61.1 & 0.031 & 0.487 [0.459, 0.516] & 0.795 [0.770, 0.814] & 0.309 \\
  & & Spec B. & 0.086 & 0.241 [0.225, 0.264] & 0.830 & 66.3 & 0.038 & 0.257 [0.237, 0.290] & 0.740 [0.683, 0.780] & 0.482 \\
\cmidrule{2-11}
  & \multirow{2}{*}{Stimuli} & Pitch & 0.086 & 0.490 [0.472, 0.505] & 0.969 & 85.3 & 0.027 & 0.991 [0.987, 0.993] & 0.993 [0.990, 0.995] & 0.002 \\
  & & BPM & 0.087 & 0.621 [0.561, 0.674] & 0.995 & 83.4 & 0.032 & 0.997 [0.994, 0.998] & 0.991 [0.988, 0.994] & $-$0.005 \\
\cmidrule{2-11}
  & \multirow{4}{*}{Vocals} & Pitch & 0.089 & 0.173 [0.166, 0.182] & 0.411 & 60.7 & 0.033 & 0.045 [0.038, 0.051] & 0.115 [0.104, 0.127] & 0.070 \\
  & & BPM & 0.087 & 0.294 [0.279, 0.310] & 0.522 & 82.3 & 0.032 & 0.182 [0.176, 0.190] & 0.289 [0.282, 0.297] & 0.107 \\
  & & Spec C. & 0.088 & 0.179 [0.167, 0.191] & 0.744 & 45.7 & 0.035 & 0.547 [0.532, 0.564] & 0.661 [0.641, 0.682] & 0.114 \\
  & & Spec B. & 0.088 & 0.245 [0.232, 0.258] & 0.718 & 76.4 & 0.032 & 0.456 [0.435, 0.480] & 0.589 [0.563, 0.616] & 0.134 \\
\midrule
\multirow{13}{*}{Vocals}
  & \multirow{4}{*}{Strings} & Pitch & 0.088 & 0.243 [0.223, 0.262] & 0.717 & 60.7 & 0.023 & 0.529 [0.510, 0.541] & 0.688 [0.661, 0.701] & 0.159 \\
  & & BPM & 0.087 & 0.157 [0.153, 0.162] & 0.531 & 48.8 & 0.032 & 0.439 [0.416, 0.460] & 0.541 [0.516, 0.564] & 0.102 \\
  & & Spec C. & 0.088 & 0.396 [0.365, 0.426] & 0.901 & 79.2 & 0.029 & 0.770 [0.762, 0.779] & 0.852 [0.846, 0.859] & 0.082 \\
  & & Spec B. & 0.089 & 0.294 [0.272, 0.318] & 0.806 & 72.6 & 0.031 & 0.648 [0.618, 0.673] & 0.758 [0.742, 0.773] & 0.110 \\
\cmidrule{2-11}
  & \multirow{3}{*}{Drums} & BPM & 0.087 & 0.207 [0.198, 0.215] & 0.473 & 70.3 & 0.021 & $-$0.022 [$-$0.051, 0.007] & 0.241 [0.197, 0.288] & 0.263 \\
  & & Spec C. & 0.091 & 0.244 [0.227, 0.269] & 0.849 & 60.8 & 0.038 & 0.661 [0.649, 0.671] & 0.814 [0.802, 0.820] & 0.153 \\
  & & Spec B. & 0.086 & 0.266 [0.243, 0.300] & 0.864 & 62.0 & 0.023 & 0.195 [0.145, 0.232] & 0.528 [0.482, 0.591] & 0.333 \\
\cmidrule{2-11}
  & \multirow{2}{*}{Stimuli} & Pitch & 0.089 & 0.389 [0.371, 0.411] & 0.957 & 79.9 & 0.030 & 0.998 [0.997, 0.999] & 0.998 [0.997, 0.999] & $-$0.000 \\
  & & BPM & 0.087 & 0.728 [0.710, 0.744] & 0.998 & 92.1 & 0.031 & 0.999 [0.999, 0.999] & 0.998 [0.997, 0.998] & $-$0.001 \\
\cmidrule{2-11}
  & \multirow{4}{*}{Vocals} & Pitch & 0.088 & 0.148 [0.138, 0.159] & 0.514 & 42.5 & 0.037 & 0.172 [0.158, 0.182] & 0.233 [0.212, 0.247] & 0.061 \\
  & & BPM & 0.088 & 0.232 [0.218, 0.245] & 0.574 & 65.9 & 0.027 & 0.412 [0.384, 0.440] & 0.575 [0.526, 0.614] & 0.162 \\
  & & Spec C. & 0.088 & 0.172 [0.152, 0.194] & 0.790 & 46.0 & 0.028 & 0.574 [0.569, 0.578] & 0.695 [0.684, 0.702] & 0.122 \\
  & & Spec B. & 0.088 & 0.220 [0.205, 0.239] & 0.801 & 66.3 & 0.024 & 0.480 [0.474, 0.485] & 0.665 [0.656, 0.672] & 0.185 \\
\midrule
\multirow{13}{*}{EnCodec}
  & \multirow{4}{*}{Strings} & Pitch & 0.088 & 0.310 [0.294, 0.325] & 0.764 & 75.4 & 0.025 & 0.697 [0.652, 0.724] & 0.709 [0.642, 0.759] & 0.013 \\
  & & BPM & 0.087 & 0.174 [0.165, 0.182] & 0.592 & 50.8 & 0.036 & 0.520 [0.472, 0.555] & 0.610 [0.548, 0.661] & 0.090 \\
  & & Spec C. & 0.088 & 0.511 [0.474, 0.540] & 0.893 & 87.4 & 0.028 & 0.867 [0.846, 0.884] & 0.899 [0.877, 0.920] & 0.032 \\
  & & Spec B. & 0.090 & 0.332 [0.305, 0.360] & 0.827 & 75.1 & 0.026 & 0.775 [0.735, 0.805] & 0.840 [0.789, 0.879] & 0.065 \\
\cmidrule{2-11}
  & \multirow{3}{*}{Drums} & BPM & 0.087 & 0.181 [0.161, 0.207] & 0.522 & 51.5 & 0.025 & 0.448 [0.393, 0.505] & 0.567 [0.459, 0.642] & 0.119 \\
  & & Spec C. & 0.089 & 0.269 [0.236, 0.309] & 0.835 & 61.6 & 0.034 & 0.838 [0.817, 0.855] & 0.872 [0.837, 0.894] & 0.034 \\
  & & Spec B. & 0.087 & 0.260 [0.225, 0.300] & 0.812 & 58.3 & 0.032 & 0.820 [0.792, 0.834] & 0.858 [0.846, 0.868] & 0.038 \\
\cmidrule{2-11}
  & \multirow{2}{*}{Stimuli} & Pitch & 0.088 & 0.564 [0.508, 0.625] & 0.988 & 85.7 & 0.028 & 0.994 [0.991, 0.995] & 0.991 [0.988, 0.993] & $-$0.003 \\
  & & BPM & 0.089 & 0.661 [0.614, 0.699] & 0.997 & 94.0 & 0.030 & 0.999 [0.998, 0.999] & 0.994 [0.988, 0.997] & $-$0.005 \\
\cmidrule{2-11}
  & \multirow{4}{*}{Vocals} & Pitch & 0.088 & 0.175 [0.166, 0.185] & 0.468 & 53.8 & 0.030 & 0.189 [0.172, 0.203] & 0.185 [0.162, 0.207] & $-$0.004 \\
  & & BPM & 0.088 & 0.237 [0.216, 0.260] & 0.519 & 64.5 & 0.022 & 0.418 [0.359, 0.457] & 0.579 [0.479, 0.647] & 0.161 \\
  & & Spec C. & 0.088 & 0.202 [0.177, 0.241] & 0.789 & 48.6 & 0.031 & 0.789 [0.748, 0.817] & 0.778 [0.727, 0.814] & $-$0.011 \\
  & & Spec B. & 0.088 & 0.197 [0.180, 0.212] & 0.684 & 57.4 & 0.023 & 0.767 [0.724, 0.802] & 0.759 [0.702, 0.801] & $-$0.008 \\
\bottomrule
\end{tabular}
\end{table}

\clearpage

\subsection{Comparison Tests}
\label{app_comparison}

\begin{table*}[!h]
\centering
\caption{\textbf{Natural versus synthetic encoding across architectures.} Layer-matched Wilcoxon signed-rank tests (unit $=$ layer pair, $N$ pairs) for RAVE and a pretrained EnCodec decoder. Nat and Syn give medians [IQR]. $\Delta$ (HL) is the Hodges--Lehmann synthetic$-$natural difference with 95\% CI (positive $=$ synthetic higher); $r$ is rank-biserial. $p_{\text{adj}}$ is the Benjamini--Hochberg-adjusted exact sign-permutation $p$ within each measure family. Spectral features excluded (no synthetic spectral analysis).}
\label{tab:supp_nat_syn}
\renewcommand{\arraystretch}{1.15}\setlength{\tabcolsep}{4pt}\footnotesize
\begin{tabular}{l l l r r r r r r r }
\toprule
\textbf{Arch.} & \textbf{Feature} & \textbf{Measure} & $N$ & \textbf{Nat} & \textbf{Syn} & $\Delta$ (HL) [95\% CI] & $r$ & ICC & $p_{\text{adj}}$ \\
\midrule
\multirow{6}{*}{RAVE}
& \multirow{3}{*}{Pitch}
  & Mean $|\rho|$ & 84 & 0.199 & 0.436 & +0.248 [+0.232, +0.263] & +1.00 & 0.450 & $<$.001~*** \\
& & \% $>$ p95 & 84 & 74.6 & 90.1 & +15.76 [+14.35, +17.19] & +1.00 & 0.507 & $<$.001~*** \\
& & $R^2$ & 84 & 0.365 & 0.997 & +0.612 [+0.594, +0.630] & +1.00 & 0.765 & $<$.001~*** \\
\cmidrule{2-10}
& \multirow{3}{*}{BPM}
  & Mean $|\rho|$ & 84 & 0.197 & 0.752 & +0.562 [+0.524, +0.625] & +1.00 & 0.835 & $<$.001~*** \\
& & \% $>$ p95 & 84 & 77.8 & 97.4 & +18.27 [+16.67, +19.88] & +1.00 & 0.754 & $<$.001~*** \\
& & $R^2$ & 84 & 0.432 & 0.996 & +0.566 [+0.551, +0.580] & +1.00 & 0.097 & $<$.001~*** \\
\midrule
\multirow{6}{*}{EnCodec}
& \multirow{3}{*}{Pitch}
  & Mean $|\rho|$ & 13 & 0.244 & 0.555 & +0.319 [+0.256, +0.381] & +1.00 & --- & $<$.001~*** \\
& & \% $>$ p95 & 13 & 81.6 & 91.0 & +12.11 [+9.38, +14.65] & +1.00 & --- & $<$.001~*** \\
& & $R^2$ & 13 & 0.478 & 0.994 & +0.537 [+0.499, +0.582] & +1.00 & --- & $<$.001~*** \\
\cmidrule{2-10}
& \multirow{3}{*}{BPM}
  & Mean $|\rho|$ & 13 & 0.185 & 0.649 & +0.470 [+0.401, +0.541] & +1.00 & --- & $<$.001~*** \\
& & \% $>$ p95 & 13 & 75.5 & 96.5 & +19.12 [+14.91, +23.77] & +1.00 & --- & $<$.001~*** \\
& & $R^2$ & 13 & 0.647 & 0.998 & +0.388 [+0.322, +0.484] & +1.00 & --- & $<$.001~*** \\
\bottomrule
\end{tabular}
\end{table*} \begin{table}[!h]
\centering
\caption{\textbf{Depth curvature, EnCodec (descriptive).} Median per-cell quadratic coefficient $\beta_2$ over normalized depth (positive $=$ U-shaped, negative $=$ inverted-U) with [IQR] and rank-biserial $r$. As EnCodec is a single model, each feature contributes only $N{=}2$--$3$ cells; the across-cell Wilcoxon cannot reach significance ($p\geq0.25$ throughout), so results are descriptive.}
\label{tab:supp_depth_encodec}
\renewcommand{\arraystretch}{1.15}\setlength{\tabcolsep}{4pt}\footnotesize
\begin{tabular}{l l r r r}
\toprule
\textbf{Measure} & \textbf{Feature} & $N$ & $\beta_2$ [IQR] & $r$ \\
\midrule
\multirow{4}{*}{Mean $|\rho|$}
  & Pitch               & 2 & -0.14 [-0.16, -0.12] & -1.00 \\
  & BPM                 & 3 & 0.16 [0.08, 0.21]    & 0.67  \\
  & Spectral Centroid   & 3 & 0.29 [0.03, 0.30]    & 0.67  \\
  & Spectral Bandwidth  & 3 & 0.15 [0.10, 0.20]    & 1.00  \\
\midrule
\multirow{4}{*}{\% $>$ p95}
  & Pitch               & 2 & -26.59 [-33.70, -19.49] & -1.00 \\
  & BPM                 & 3 & 19.72 [6.30, 40.94]     & 0.67  \\
  & Spectral Centroid   & 3 & 10.28 [-3.09, 31.14]    & 0.33  \\
  & Spectral Bandwidth  & 3 & 0.60 [-9.86, 23.86]      & 0.33  \\
\midrule
\multirow{4}{*}{$R^2$}
  & Pitch               & 2 & -0.48 [-0.56, -0.39] & -1.00 \\
  & BPM                 & 3 & -0.78 [-0.85, -0.59] & -1.00 \\
  & Spectral Centroid   & 3 & -0.31 [-0.42, -0.29] & -1.00 \\
  & Spectral Bandwidth  & 3 & -0.60 [-0.62, -0.39] & -1.00 \\
\bottomrule
\end{tabular}
\end{table} 
\begin{figure*}
    \centering
    \includegraphics[width=1.0\linewidth]{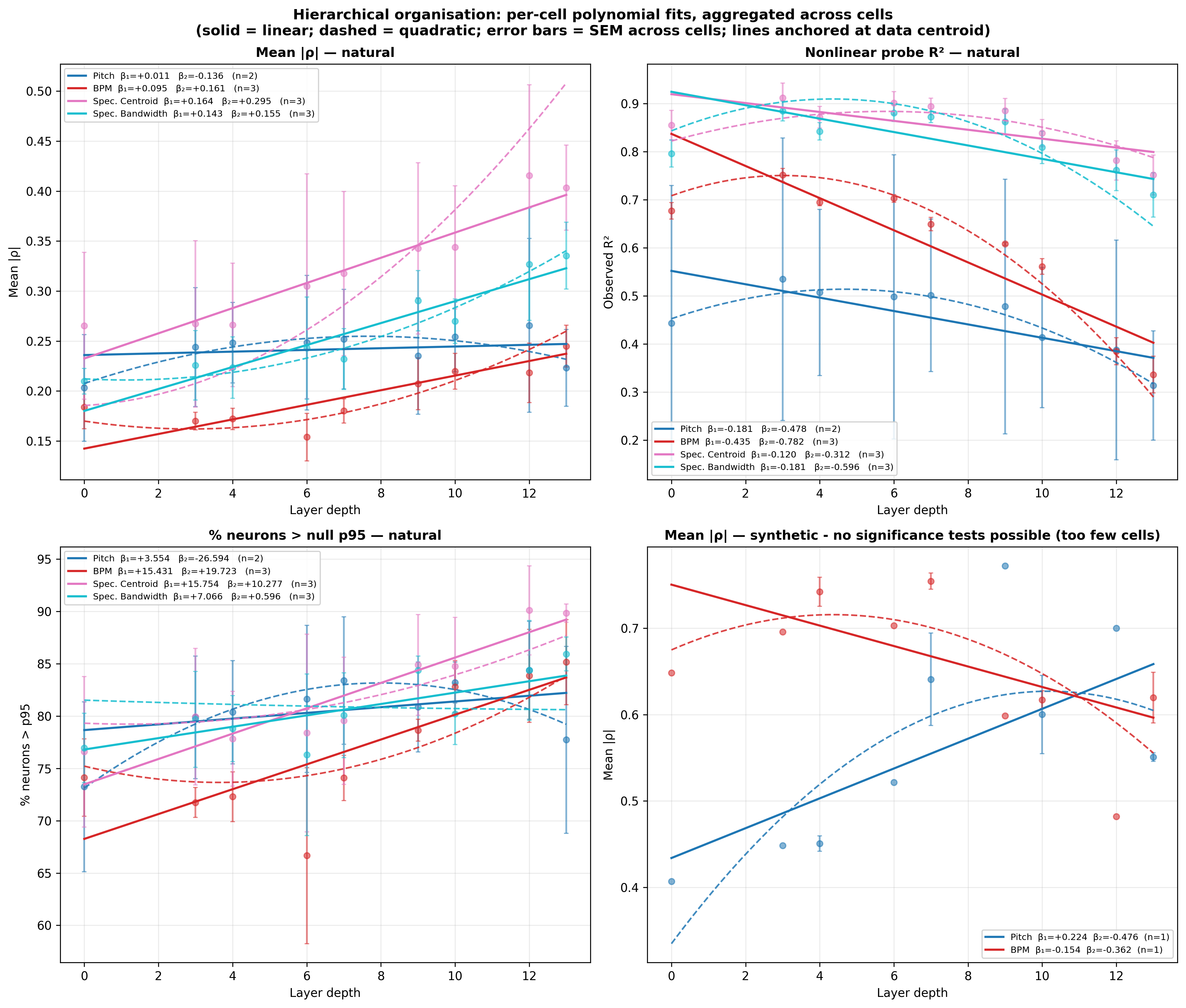}
    \caption{Plots of per layer mean $|\rho|$, neuron prevalence and $R^2$ across four audio features for natural audio with the EnCodec model. Only per layer mean $|\rho|$ displayed for synthetic audio as other features are near ceiling throughout (e.g.\ flat). Significance tests not possible as N=3 for natural and N=1 for synthetic.}
    \label{fig:depth_encodec}
\end{figure*}

\begin{table*}[t]
\centering
\caption{\textbf{Cross-layer cluster versus best single layer, natural audio only.} Hodges--Lehmann estimate of the median per-cell difference (best cross-layer cluster $-$ best single layer within a section) with 95\% bootstrap CI, for RAVE and a pretrained EnCodec decoder; positive values indicate the cross-layer cluster outperforms the best single layer. Layer/cluster mean are the raw (non-paired) means; best cluster/layer $n$ are the average number of units in the selected cluster vs.\ the best single layer. $r$ is the matched-pairs rank-biserial effect size; $p_{\text{adj}}$ is the Benjamini--Hochberg-adjusted exact sign-permutation $p$ within each measure family. Between-model (mod) and between-dataset (dat) intraclass correlations index dependence among cells. EnCodec is a single model, so between-model ICC is undefined (--).}
\label{tab:cluster_combined}
\renewcommand{\arraystretch}{1.25}\setlength{\tabcolsep}{3pt}\scriptsize
\begin{tabular}{l l l r r r r r r r r r r}
\toprule
& & & & & & & & \multicolumn{2}{c}{ICC} & & \multicolumn{2}{c}{Avg. best $n$} \\
\cmidrule(lr){9-10}\cmidrule(lr){12-13}
\textbf{Arch.} & \textbf{Meas.} & \textbf{Feat.} & $N$ & Layer & Cluster & \textbf{HL [95\% CI]} & $r$ & mod & dat & $p_{\text{adj}}$ & cluster & layer \\
\midrule
\multirow{12}{*}{RAVE}
& \multirow{4}{*}{Mean $|\rho|$}
  & Pitch              & 18 & 0.239 & 0.237 & $-$0.004 [$-$0.015, +0.012]   & $-$0.05 & 0.154 & 0.339 & 0.769 & 779 & 480 \\
& & BPM                 & 27 & 0.233 & 0.257 & +0.019 [+0.007, +0.033]       & +0.65 & 0.000 & 0.411 & 0.006** & 360 & 309 \\
& & Spec. Centroid      & 27 & 0.377 & 0.362 & $-$0.018 [$-$0.036, +0.009]   & $-$0.29 & 0.268 & 0.008 & 0.400 & 1102 & 513 \\
& & Spec. Bandwidth     & 27 & 0.348 & 0.335 & $-$0.022 [$-$0.044, +0.004]   & $-$0.34 & 0.000 & 0.000 & 0.605 & 831 & 466 \\
\cmidrule{2-13}
& \multirow{4}{*}{\% $>$ p95}
  & Pitch              & 18 & 82.6 & 86.9 & +3.91 [+0.13, +8.37]          & +0.50 & 0.183 & 0.153 & 0.061 & 514 & 420 \\
& & BPM                 & 27 & 84.9 & 91.0 & +6.15 [+3.23, +9.07]          & +0.75 & 0.000 & 0.233 & 0.001** & 486 & 309 \\
& & Spec. Centroid      & 27 & 88.9 & 88.0 & $-$0.59 [$-$3.17, +1.53]      & $-$0.13 & 0.137 & 0.142 & 0.521 & 971 & 442 \\
& & Spec. Bandwidth     & 27 & 89.3 & 89.8 & $-$0.02 [$-$2.45, +3.18]      & $-$0.01 & 0.014 & 0.083 & 0.686 & 738 & 423 \\
\cmidrule{2-13}
& \multirow{4}{*}{$R^2$}
  & Pitch              & 18 & 0.412 & 0.418 & +0.005 [+0.001, +0.010]       & +0.56 & 0.000 & 0.000 & 0.083 & 738 & 352 \\
& & BPM                 & 27 & 0.490 & 0.500 & +0.011 [$-$0.009, +0.036]      & +0.17 & 0.000 & 0.000 & 0.860 & 953 & 338 \\
& & Spec. Centroid      & 27 & 0.788 & 0.789 & +0.001 [$-$0.008, +0.009]     & +0.02 & 0.000 & 0.004 & 0.917 & 825 & 359 \\
& & Spec. Bandwidth     & 27 & 0.744 & 0.733 & $-$0.004 [$-$0.015, +0.012]   & $-$0.15 & 0.002 & 0.093 & 0.860 & 698 & 352 \\
\midrule
\multirow{12}{*}{EnCodec}
& \multirow{4}{*}{Mean $|\rho|$}
  & Pitch              & 6  & 0.273 & 0.355 & +0.066 [+0.018, +0.156]       & +1.00 & --- & 0.556 & 0.125 & 55 & 235 \\
& & BPM                 & 9  & 0.228 & 0.227 & $-$0.002 [$-$0.016, +0.014]   & $-$0.11 & --- & 0.000 & 0.840 & 396 & 331 \\
& & Spec. Centroid      & 9  & 0.369 & 0.436 & +0.046 [$-$0.002, +0.132]     & +0.60 & --- & 0.667 & 0.164 & 122 & 288 \\
& & Spec. Bandwidth     & 9  & 0.299 & 0.336 & +0.035 [$-$0.012, +0.079]     & +0.51 & --- & 0.070 & 0.167 & 259 & 188 \\
\cmidrule{2-13}
& \multirow{4}{*}{\% $>$ p95}
  & Pitch              & 6  & 86.9 & 92.6 & +5.66 [+3.66, +7.62]          & +1.00 & --- & 0.000 & 0.042* & 55 & 219 \\
& & BPM                 & 9  & 83.9 & 84.5 & +0.18 [$-$4.37, +6.55]        & +0.06 & --- & 0.139 & 0.859 & 207 & 295 \\
& & Spec. Centroid      & 9  & 87.3 & 90.5 & +3.28 [+0.96, +5.60]          & +0.82 & --- & 0.290 & 0.042* & 79 & 228 \\
& & Spec. Bandwidth     & 9  & 86.3 & 91.4 & +5.69 [+0.71, +8.77]          & +0.82 & --- & 0.000 & 0.042* & 211 & 203 \\
\cmidrule{2-13}
& \multirow{4}{*}{$R^2$}
  & Pitch              & 6  & 0.479 & 0.480 & $-$0.002 [$-$0.019, +0.024]   & $-$0.14 & --- & 0.138 & 0.938 & 243 & 277 \\
& & BPM                 & 9  & 0.615 & 0.629 & +0.012 [$-$0.008, +0.038]     & +0.33 & --- & 0.000 & 0.344 & 309 & 277 \\
& & Spec. Centroid      & 9  & 0.868 & 0.851 & $-$0.014 [$-$0.050, +0.021]   & $-$0.51 & --- & 0.664 & 0.344 & 365 & 277 \\
& & Spec. Bandwidth     & 9  & 0.843 & 0.777 & $-$0.065 [$-$0.117, $-$0.014] & $-$0.73 & --- & 0.623 & 0.125 & 378 & 277 \\
\bottomrule
\end{tabular}
\end{table*} 

\clearpage

\subsection{Robustness to $k$}
\label{app_robustness}
\begin{figure}[!h]
  \includegraphics[width=0.9\textwidth]{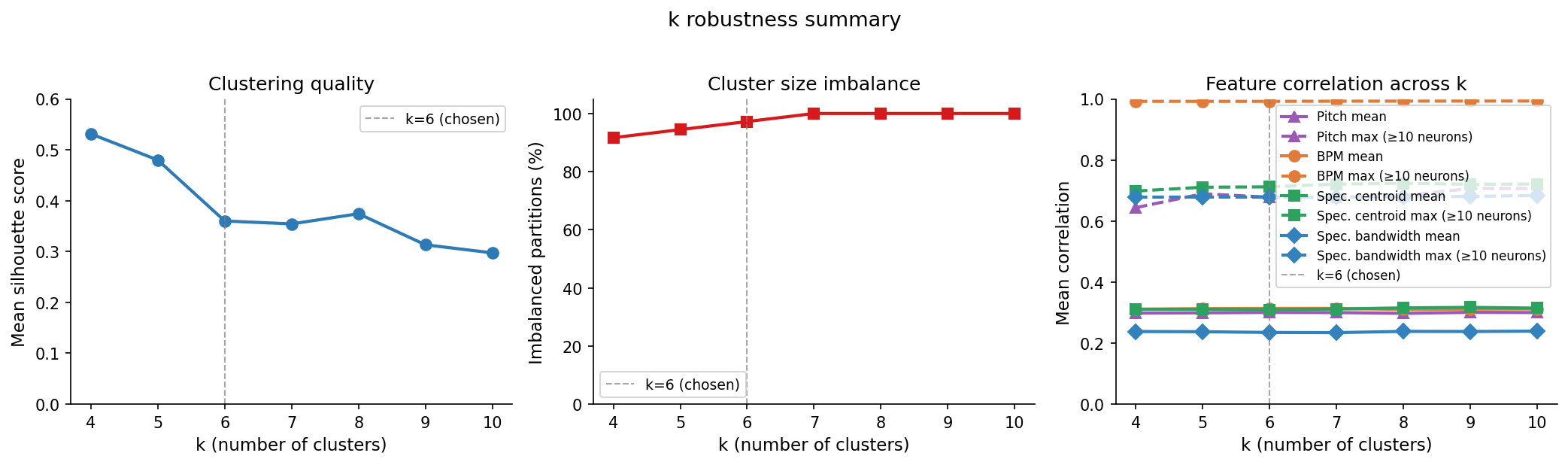}
  \caption{Plots of mean silhouette score, cluster imbalance and mean and max feature correlation across values of $k$.}
\end{figure}
\begin{figure}[!h]
  \includegraphics[width=0.9\textwidth]{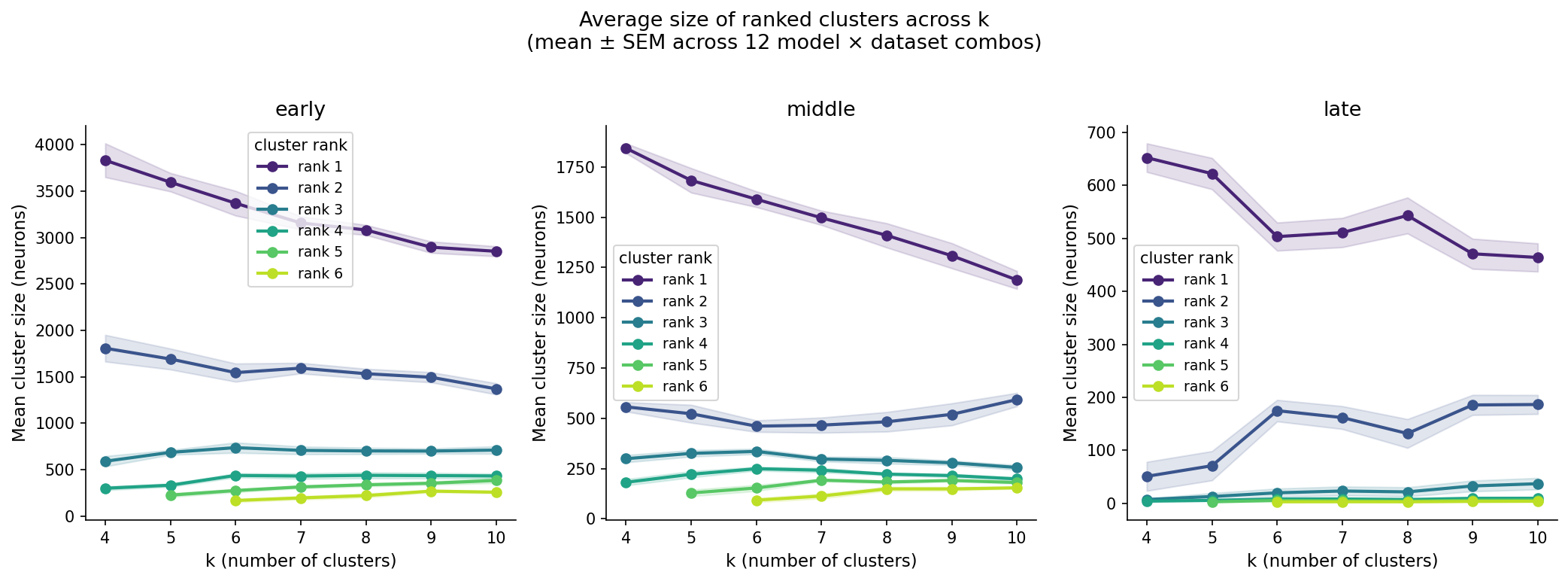}
  \caption{Plot of mean cluster size (ranked by size) across values of $k$.}
\end{figure}
\begin{figure}[!h]
  \includegraphics[width=0.9\textwidth]{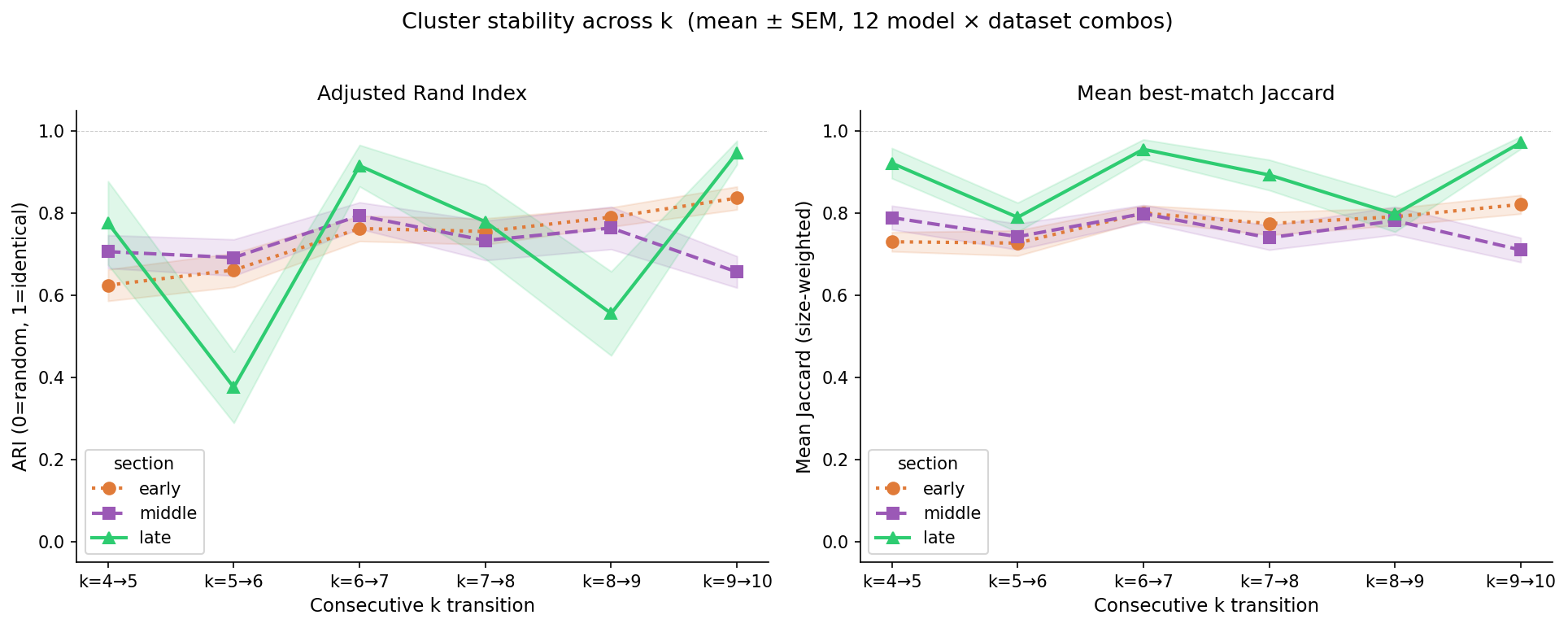}
  \caption{Plot of mean ARI for all clusters, and mean Jaccard for best cluster across values of $k$.}
\end{figure}